\documentclass[
  aps,
  prx,
  superscriptaddress,
  floatfix,
  longbibliography,
  reprint,
  hidelinks
]{revtex4-2}

\usepackage{newtxtext,newtxmath, bm}
\usepackage{graphicx}
\usepackage{siunitx, dcolumn, xcolor, url}

\begin{document}
\title{ A quantum-coherent photon--emitter interface in the original telecom band }

\author{
    Marcus Albrechtsen$^{1\ast\dagger}$,
    Severin Krüger$^{2\dagger}$,
    Juan Loredo$^{3}$,
    Lucio Stefan$^{3}$,
    Zhe Liu$^{1}$,
    Yu Meng$^{1}$,
    Lukas L.\ Niekamp$^{4}$,
    Bianca F.\ Seyschab$^{4}$,
    Nikolai Spitzer$^{2}$,
    Richard J.\ Warburton$^{4}$,
    Peter Lodahl$^{1,3}$,
    Arne Ludwig$^{2}$,
    Leonardo Midolo$^{1\ast}$
    \\[1ex]
    {\small \textit{
    $^{1}$Center for Hybrid Quantum Networks (Hy-Q), Niels Bohr Institute, University of Copenhagen,\\
    \hspace{1em}Jagtvej 155A, DK-2200 Copenhagen, Denmark.\\
    $^{2}$Lehrstuhl für Angewandte Festkörperphysik, Ruhr-Universität Bochum,\\
    \hspace{1em}Universitätsstraße 150, 44801 Bochum, Germany.\\
    $^{3}$Sparrow Quantum ApS, Nordre Fasanvej 215, DK-2000 Frederiksberg, Copenhagen, Denmark.\\
    $^{4}$Department of Physics, University of Basel, Klingelbergstrasse 82, CH-4056 Basel, Switzerland.}\\
    $^\ast$Corresponding authors. E-mails: m.albrechtsen@nbi.ku.dk, midolo@nbi.ku.dk\\
    $^\dagger$These authors contributed equally.
    }
}

\date{\today}

\begin{abstract}
Quantum dots stand out as the most advanced and versatile light-matter interface available today. Their ability to deliver high-quality, high-rate, and pure photons has set benchmarks that far surpass other emitters. Yet, a critical frontier has remained elusive: achieving these exceptional capabilities at telecom wavelengths, bridging the gap to fiber-optic infrastructure and scalable silicon photonics. Overcoming this challenge demands high-quality quantum materials and devices which, despite extensive efforts, have not been realized yet. Here, we demonstrate waveguide-integrated quantum dots and realize a fully quantum-coherent photon-emitter interface operating in the original telecommunication band. The quality is assessed by recording transform-limited linewidths only 8~\% broader than the inverse lifetime and bright 41.7~MHz emission rate under 80~MHz $\pi$-pulse excitation, unlocking the full potential of quantum dots for scalable quantum networks.
\end{abstract}

\maketitle

Photonics offers a unique opportunity in quantum technology: as the sole viable approach for interconnecting distant quantum systems, it enables the realization of scalable quantum networks by encoding qubits in telecom-wavelength photons \cite{wang2025scalable,yin2017satellite,wang2020integrated}. Secure digital data encryption can be realized using quantum key distribution (QKD) protocols, which---in combination with coherent single-photon sources---provide a pathway to scalable hardware for the quantum internet \cite{lu2021quantum,uppu2021Quantum‑dot‑based}, such as quantum repeaters \cite{azuma2015all,borregaard2020one} and device-independent QKD (DI-QKD) \cite{kolodynski2020device}. Furthermore, scalable, high-fidelity, and high-rate photon sources are the key missing-link for large-scale photonic fault-tolerant quantum computing (FTQC)  \cite{psiquantum2025manufacturable}, and deterministic photon-emitter interfaces provide a highly promising approach \cite{chan2025practical} that is intimately connected to quantum materials growth and device development constituting a growing area of interdisciplinary scientific exploration \cite{de2021materials}.

Self-assembled indium arsenide (InAs) quantum dots (QDs) are among the most mature and well-studied quantum emitters \cite{lodahl2015Interfacing}, known for their unmatched quality in terms of deterministic light-matter interface, emission purity, narrow linewidths, indistinguishability and efficiency \cite{uppu2020scalable,tomm2021bright,ding2025high}. Yet, a major long-standing hurdle to the development of a scalable quantum technology based on QDs is their operation at the near-infrared $~$930~nm wavelength, which has so far made these emitters incompatible both with standard low-loss silica fibre optics for long-distance communication as well as with silicon photonic integrated circuits. Efforts to push the emission wavelengths of QDs to the telecommunication bands \cite{alloing2005growth,holewa2025solid,nawrath2023bright,joos2024coherently,srocka2020deterministically} have so far resulted in low efficiency, high blinking, incoherent emission and poor indistinguishability, which precludes their adoption in advanced quantum applications such as measurement-based FTQC, DI-QKD and more. Similarly, alternative solid-state telecom emitters, such as T-/G-centers in Si \cite{komza2024indistinguishable,simmons2024scalable}, erbium ions \cite{ourari2023indistinguishable} and 2D materials \cite{zhao2021site}, while presenting other advantages such as long-lived quantum memories, fall prohibitively short in emission efficiency and quantum coherence due to uncontrolled noise processes and non-radiative decay channels combined with very long radiative lifetimes. 

In every photonic quantum-information protocol, coherence is the key element to achieve high-fidelity operation: it is relevant for applications that require multi-photon interference as well as for non-linear light-matter interactions. In quantum emitters, coherence is directly revealed by measuring the linewidth of the optical transitions, thereby probing the stability across hundreds of excitation-emission cycles \cite{uppu2020scalable,ding2025high}. The optical linewidth is broadened by all noise processes, from pure phonon-induced dephasing to slow spectral diffusion. Transform-limited linewidths, i.e., transitions limited by the inverse lifetime ($\Gamma\simeq(2\pi\tau)^{-1}$), therefore enable a high coherent photon-emitter cooperativity when realized in nanophotonic devices such as waveguides and cavities \cite{borregaard2019quantum,kuhlmann2015transform}. While photonic nanostructures can strongly enhance the emission into the mode via Purcell enhancement \cite{manga2007single},  non-radiative recombination, dephasing rates and spectral diffusion are fundamentally rooted in the material design, quality, and processing, which is where near-infrared QDs traditionally excel.

Here, we report for the first time self-assembled InAs QDs emitting in the original telecommunication band (or O-band, 1260--1360 nm), with characteristics matching their 930-nm-wavelength counterparts both in terms of quantum coherence, near-lifetime limited optical linewidths, and efficiency. We developed a state-of-the-art molecular-beam epitaxial growth process of high-quality QDs in suspended GaAs waveguides and employ $p$-$i$-$n$ diode junctions to suppress charge noise. The native integration of our light-matter interface with waveguides directly enables complex photonic integrated circuits including compatibility with silicon photonics.

\begin{figure*}
    \centering
    \includegraphics[width=0.9\textwidth]{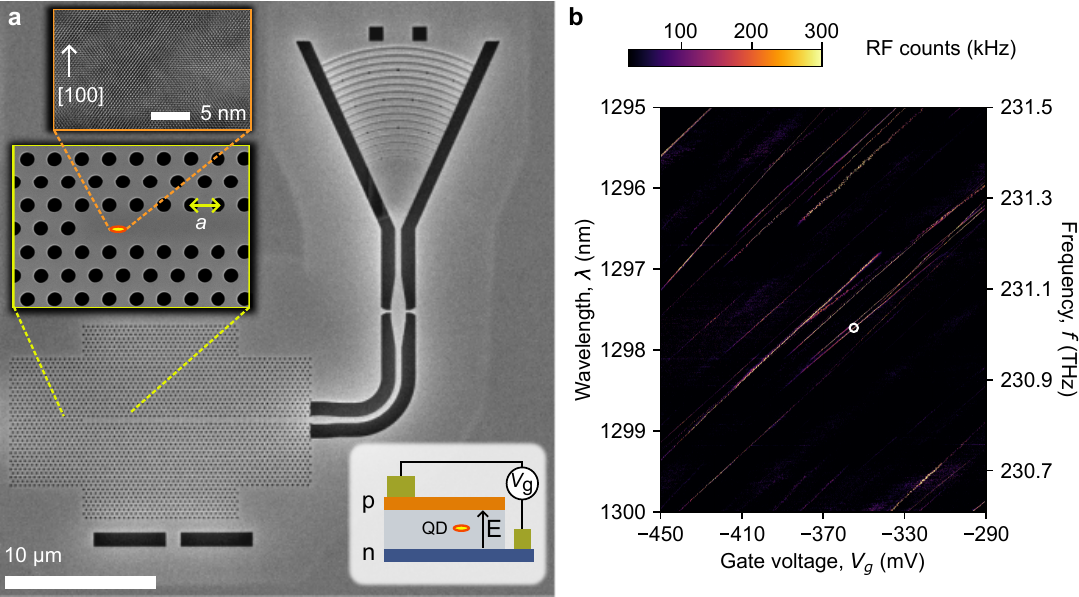}
    \caption{
        \textbf{Telecom quantum dots in a gated photonic crystal waveguide.}
        \textbf{a}, Scanning electron micrograph of a photonic nanostructure with lattice constant, $a=\SI{338}{nm}$, hosting gated telecom quantum dots. Inset: Scanning transmission electron microscope  cross-section of the GaAs membrane in the QD area, showing defect-free epitaxial growth.  
        \textbf{b}, Resonance fluorescence (RF) map under continuous wave excitation. The white circle at ($-\SI{355}{mV}$, 1297.726~nm, 231.013~THz) indicates the main resonance considered.
    }
    \label{fig:1}
\end{figure*}

\section*{Results}
\subsection*{Quantum-coherent single-photon devices}

To shift the wavelength to \SI{1.3}{\micro\m} while retaining ultra-low noise, we developed high-quality pseudomorphic growth that avoids dislocations, which are known sources of charge noise and dark decay channels. This was done by embedding the InAs QDs in a 7~nm In$_{0.3}$Ga$_{0.7}$As quantum well, which relaxes the vertical strain across the QDs. This makes the QDs taller than standard QDs, shifting their fundamental transition-energy to approximately 1~eV.
The larger QD size additionally benefits their use as spin-photon interfaces since the hyperfine coupling of the electron spin to the nuclear bath is reduced \cite{warburton2013single}.
An important correlation between high optical brightness and low surface roughness was found during the material growth development process, which directly enabled the optimized process used for our devices. Moreover, we combine high-quality growth with electrical control of QDs to suppress charge noise \cite{kuhlmann2013charge}. Details about the growth process, the sample design and fabrication are provided in the Methods section.

Figure~\ref{fig:1}\textbf{a} shows the device used in the experiment, consisting of a photonic crystal waveguide interfaced to a suspended nanobeam \cite{arcari2014near} and a focusing grating coupler \cite{zhou2018high} designed for operation at 1310 nm under cryogenic conditions.
The self-assembled QDs are embedded in the suspended GaAs membranes with electrical contacts to control the emission wavelength via Stark-tuning.
The images show the dislocation-free crystal growth and high-quality sample fabrication, and we extract a low propagation loss of $(3.5\pm0.7)$~dB/mm in the doped nanobeam waveguide, which corresponds to only 100 mdB in our \SI{30}{\micro\m} circuit (see Methods).
Since telecom QDs are spectrally much farther from the GaAs bandgap edge compared to the standard 930~nm QDs, electro-absorption \cite{wang2021electroabsorption} is highly reduced, and allows operation at low voltages and low currents, compatible with complementary metal-oxide semiconductor (CMOS) levels.

\begin{figure*}
    \centering
    \includegraphics[width=0.9\textwidth]{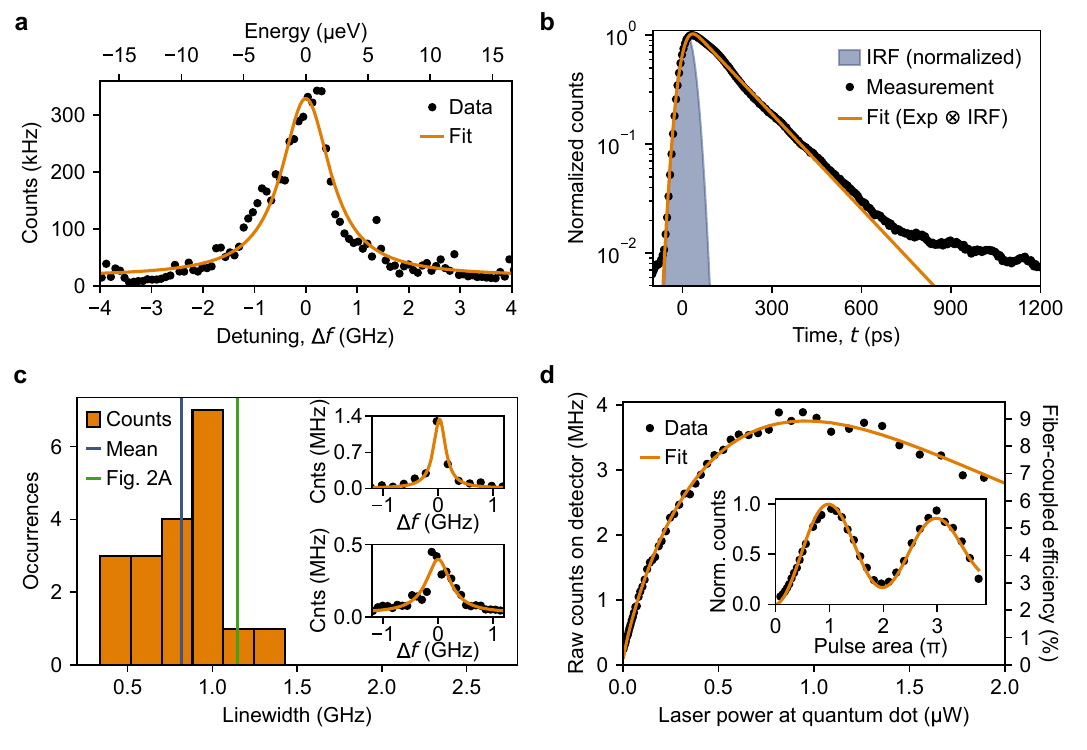}
    \caption{
        \textbf{Coherent single-photon emission from telecom quantum dots.}
        \textbf{a}, Resonance fluorescence of a quantum dot (neutral exciton) featuring a transform-limited linewidth. The fit is a Lorentzian with full-width half-maximum of $\Gamma_{RF}$ = $1.15(5)$~GHz. \textbf{b}, Time-resolved fluorescence of the emission line in A, with lifetime $\tau = 150(2)$~ps, corresponding to a Fourier-transformed-limited linewidth of $\gamma_1=1.08(1)$~GHz. A slow background, which contributes to $\sim1$\% of the total count rate, with a decay time of $3.1(1)$~ns is observed, likely stemming from background emission of neighboring QDs. 
        \textbf{c}, Statistical distribution of the mean linewidth of the 19 quantum dots in Fig.~1\textbf{b} with overall average linewidth 0.8~GHz. Insets: Additional RF scans with $\lambda=1297.640$~nm, $V_g=-369$~mV, and $0.26(1)$~GHz linewidth (top) and $\lambda=1297.665$~nm, $V_g=-350$~mV, and $0.53(4)$~GHz linewidth (bottom).
        \textbf{d}, Detected counts as a function of laser pulse power, resulting in a peak rate of 3.9~MHz into the detector at $\pi$-pulse. Inset: Rabi oscillations obtained with a narrow (1.5~GHz) etalon filter demonstrating coherent nature of the quantum emitter.
    }
    \label{fig:2}
\end{figure*}

We employ resonance fluorescence (RF) spectroscopy to characterize the QDs (see Methods). Figure~\ref{fig:1}\textbf{b} shows a map of the RF signal as a function of the applied bias and excitation laser wavelength, where multiple sub-GHz narrow-linewidth transitions are observed. The emission frequency can be tuned over a wide 400~GHz bandwidth at a rate of 3.5--5.0~GHz/mV, which is key to scale up the platform using multiple emitters \cite{papon2023independent,tiranov2023collective} that requires compensating the inhomogeneous broadening.

The emitter coherence is tested by comparing the RF full-width half maximum $\Gamma_\text{RF}$ of a given transition to the corresponding lifetime-limited linewidth $\gamma_1$. In the presence of a pure dephasing rate $\gamma_\text{dp}$, the corresponding RF linewidth is a Lorentzian, broadened by $\Gamma_\text{RF} = \gamma_1 + 2\gamma_\text{dp}$ \cite{lodahl2015Interfacing}. So far, RF measurements on telecom emitters have resulted in broad Gaussian (or Voigt) lineshapes, limited by spectral diffusion and bounding the coherence between individual photons.
Figure~2\textbf{a} shows a single RF scan across a neutral exciton with a fixed applied bias of $-\SI{355}{mV}$ where each data point is acquired with a 0.1~s integration time after stepping the CW laser in \SI{0.5}{pm} steps. Such a slow scan does not exhibit signature of spectral diffusion over the whole acquisition and results in a Lorentzian shape with a linewidth of $(1.15 \pm 0.05)$ GHz. To extract the natural linewidth we perform time-resolved photoluminescence of the same line with a pulsed laser (Fig.~2\textbf{b}), resulting in a lifetime $\tau = 150(2)$~ps or $\gamma = (2\pi\tau)^{-1} = (1.08 \pm 0.01)$ GHz, which immediately implies that dephasing is widely suppressed in this sample. The relatively fast decay of this QD, compared to the average of QDs tested can be explained by a moderate Purcell enhancement ($\sim5$--10) due to the slow-light effect in the PhC waveguide.

Figure~2\textbf{c} shows a histogram of the average linewidth extracted as fits across 19 transitions traced in Fig.~1\textbf{b}. We consistently observe narrow linewidths with an average of \SI{0.8}{GHz} across all lines and the two insets show two example narrow lines of \SI{0.53(4)}{GHz} and \SI{0.26(1)}{GHz} from different QDs.
Since the spontaneous emission rate depends on the location of the QD in the waveguide as well as the spectral proximity to the PhC bandedge, we observe a variation in linewidths even for the same QD.

The QDs are driven with a resonant 10-ps long $\pi$-pulse to achieve full population inversion, providing an accurate measurement of the maximum attainable photon rate. We collect the photons via the focused grating coupler and measure independently the losses of each optical component in the path to extract the collection efficiency of the single-photon source. Fig.~\ref{fig:2}\textbf{d} shows the power-dependent count-rate measured on the detector when exciting the source with a 80~MHz repetition rate. A maximum count rate of 3.9~MHz at $\pi$-area pulse is recorded at the detector (efficiency of 70~\%). Normalizing by the loss of the grating filter and detector, we estimate 7.6~MHz photon counts in the fiber.
We estimate the losses in the collection due to the grating coupler, free-space optics and fiber coupling  to be 18.2~\%, i.e.  the GaAs device produces a rate of 41.7~MHz (efficiency of 52~\%) highly coherent single photons. This is on par with the efficiencies reported for 930~nm wavelength devices \cite{tomm2021bright,uppu2020scalable,ding2025high} but widely exceeding the state-of-the-art for telecom emitters. Moreover, we observe clear Rabi oscillations with power (inset of Fig.~\ref{fig:2}\textbf{d}) which confirm the high quantum coherence of the emitter.

\begin{figure}
\centering
    \includegraphics[width=0.45\textwidth]{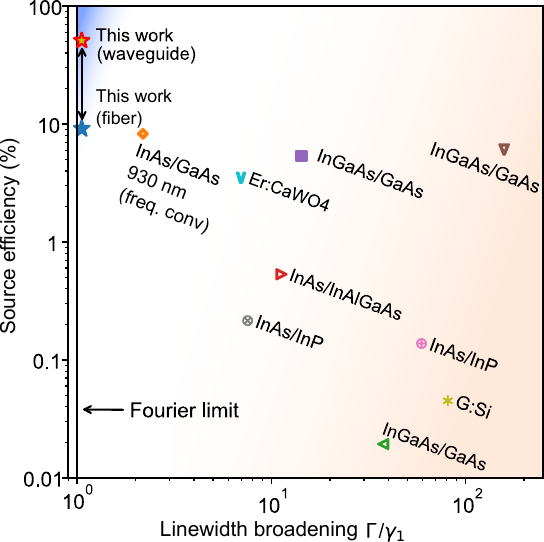}
   \caption{
        \textbf{State-of-the-art of telecom quantum emitters.}
        Values of photon collection efficiency and linewidth broadening. The data is reported from a survey of the literature, where different excitation schemes are used as well as different nanostructures. The label indicates the material platform. References: \textcolor[rgb]{0.996,0.496,0.055}{$\diamond$} \cite{zahidy2024quantum}, \textcolor[rgb]{0.172,0.625,0.172}{$\lhd$} \cite{nawrath2019coherence}, \textcolor[rgb]{0.836,0.152,0.156}{$\rhd$} \cite{hauser2025deterministic}, \textcolor[rgb]{0.578,0.402,0.738}{$\blacksquare$} \cite{joos2024coherently}, \textcolor[rgb]{0.547,0.336,0.293}{$\nabla$} \cite{nawrath2023bright}, \textcolor[rgb]{0.887,0.465,0.758}{$\oplus$} \cite{holewa2024high}, \textcolor[rgb]{0.496,0.496,0.496}{$\otimes$} \cite{srocka2020deterministically}, \textcolor[rgb]{0.734,0.738,0.133}{$\ast$} \cite{komza2024indistinguishable}, \textcolor[rgb]{0.090,0.742,0.809}{$\vee$} \cite{ourari2023indistinguishable}. Details are given in Supplementary Table~S2.}
    \label{fig:3}
\end{figure}

A comparison with other solid-state sources is provided in Fig.~\ref{fig:3}, where we plot reported linewidth broadening ($\Gamma / \gamma_1$) and source efficiency (defined as photon rate divided by repetition rate of the excitation laser) and the corresponding material system reported in the literature (for a full list of references and details, see the Supplementary Information). A perfect quantum-coherent light-matter interface exhibits Fourier transform-limited linewidths, i.e., $\Gamma/\gamma_1=1$. Once this is achieved, improving the efficiency towards unity is a photonic engineering task from enhancing the emitter-mode coupling to reducing circuit losses \cite{psiquantum2025manufacturable,aghaee2025scaling}.
This analysis clearly shows how the O-band QDs reported in this work outperform all other emitters and platforms, including standard near-infrared frequency-converted QDs \cite{da_lio2022pure}. 
The ultimate benchmark of the QD sources is the threshold for fault-tolerant photonic quantum computing. Recently a blueprint for QD sources  was analyzed \cite{chan2025practical} and thresholds established for this specific architecture tolerating approximately 10~\% loss and few percent of linewidth broadening. These demanding requirements appear to be compatible with the ultra-low-noise QD telecom platform presented here.

\subsection*{Demonstration of single-photon source operation}
A known issue in previous demonstrations of telecom solid-state emitters is the presence of strong and fast blinking and Auger processes \cite{kurzmann2016auger}. Proximity to defects causes the random escape of charge carriers, that greatly enhances non-radiative decay. Mitigating this effect is crucial for high quantum efficiency towards achieving bright single-photon sources. In the following, we characterize blinking and  multi-photon contamination in our device by recording auto-correlation histograms $g^{(2)}(\tau)$ in a Hanbury Brown and Twiss setup under pulsed resonant $\pi$-pulse excitation.

\begin{figure*}
    \centering
    \includegraphics[width=0.9\textwidth]{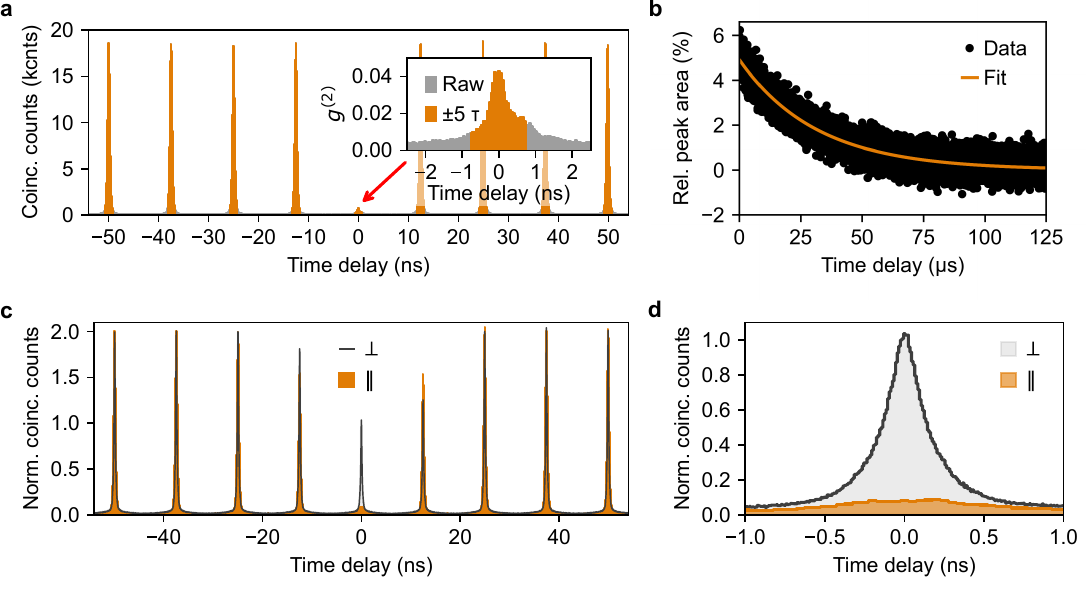}
    \caption{
        \textbf{Single-photon source operation.}
        \textbf{a}, Raw auto-correlation measurement $g^{(2)}(\tau)$ confirming single-photon emission. Inset: Zoom around the central peak in the range $\pm5\tau$, i.e., including $99.3$ \% of the collected photons. The area of the peak is normalized to that of the infinitely far-away side peaks, yielding a $g^{(2)}(0)=0.0610(2)$.
        \textbf{b}, Relative areas of side-peaks at long time-scales, showing the low-levels of blinking from a high-quality quantum emitter.
        \textbf{c}, Measurement of coincidence counts in a Hong-Ou-Mandel setup for parallel (indistinguishable) and perpendicular (fully distinguishable) configurations.
        \textbf{d}, Zoom around the central normalized peak. The raw visibility $V_\text{raw} = (83.8 \pm 0.1)$~\% is calculated over a $\pm750$ ps window, corresponding to $\sim5\tau$, i.e., including $99.3$ \% of the collected photons.
        }
    \label{fig:4}
\end{figure*}

Figure~\ref{fig:4}\textbf{a} and \ref{fig:4}\textbf{b} show the measured $g^{(2)}(\tau)$ for the QD studied in Fig.~\ref{fig:2} at short and long time-scales, respectively.  We do not observe blinking at short time-scales, crucial for applications where multiple consecutive photons are required, e.g., for a temporally de-multiplexed source \cite{sund2023high}.
By integrating the area under the first 10'000 side peaks on both sides we observe weak residual blinking as a single exponential component with just 5~\% amplitude at a \SI{31}{\micro\s} time scale. The blinking strength represents the off-on ratio of the emitter, i.e., our emitter is on during 95~\% of the time. Importantly, the blinking timescale is much longer than the radiative decay such that the blinking does not affect the ability to produce a long stream of uninterrupted back-to-back single photons. For comparison, common blinking strengths reported for telecom emitters are $>100$~\% \cite{hauser2025deterministic,komza2024indistinguishable}, which results in fundamentally bounded low quantum efficiencies $<50$\%.

The inset of Fig.~\ref{fig:4}\textbf{a} shows the raw anti-bunching at zero delay without any background subtraction and the integrated area in a 1.5~ns ($\pm5~\tau$) window, which includes $99.3$~\% of the emitted photons. This is compared to the area in the same window for far-away peaks yielding a $g^{(2)}(0)=0.061(2)$. Here, the residual counts in $g^{(2)}(\tau)$ originate primarily from leakage of the pump laser into the waveguide mode due to imperfect extinction in our setup.
Non-resonant excitation schemes such as $p$-shell and phonon assisted excitation allow trivial suppression of the pump laser through spectral filtering to unlock excellent $g^{(2)}(0)$ (applicable for simple QKD such as the BB84 protocol where coherence is not required), but the additional transitions required before the resonant decay introduce jitter, which in turn reduces the coherence across the full envelope \cite{lodahl2015Interfacing}. To illustrate this point we include in the Supplementary Information a non-resonant measurement with $g^{(2)}(0)=0.016$ where effectively all counts occur within a $\pm5~\tau$ window.

Figure~\ref{fig:4}\textbf{c} shows two-photon interference using a Hong-Ou-Mandel (HOM) apparatus with adjustable polarization control between the two paths. HOM visibility provides a way to assess the indistinguishability between consecutive photon wave-packets. Assuming a pure dephasing model (consistent with oberved Lorentzian linewidths), the indistinguishability is given by $V_\text{HOM} = \frac{\gamma_1}{\gamma_1+2\gamma_\text{dp}}$, which for the QD of Fig.~\ref{fig:2}\textbf{a}, results in $V_\text{HOM} = 0.92\pm 0.04$. We extract the visibility by comparing the area under the central peak, $A_0$, to half the area of far-away peaks, $A_\infty$, and calculate the raw visibility as $V_\text{raw}=1-(2 A_0/A_\infty)$. We obtain a lower bound of $V_\text{raw} = (83.8 \pm 0.1)\%$ over the same 1.5~ns ($\pm5\tau$) integration window.
Such raw HOM visibility far exceeds previously reported values for telecom emitters under resonance fluorescence.

Supplementary Fig.~S4 shows a resonant HOM measurement on a different QD, corroborating that the coherent nature observed is generic in the sample and due to excellent growth quality. The observed transform-limited linewidths provide strong evidence that the observed HOM indistinguishability is currently only limited by experimental instrumentation and not by fundamental dephasing processes, which have instead been fully suppressed.

Figure~\ref{fig:4}\textbf{d} illustrates the interference at short delays between the co- and cross-polarized case, and a reduction of the entire envelope can be observed.
Importantly, we do not perform post-selection of the coherent photons, e.g., by time-gating the recorded events, which is possible but would compromise the count rate, severely restricting impact and applications. The discrepancy between the measured $V_\text{raw}$ and the dephasing-induced limit $V_\text{HOM}$ is to be attributed entirely to setup imperfections, such as the leakage of two-photon component from the excitation laser, i.e., by the value of $g^{(2)}(0)$.
We note that it is possible to correct $V_\text{raw}$ for $g^{(2)}(\tau)$ to reach $V_\text{HOM}$ but this does not change the fundamental emitter properties governed by the linewidth \cite{gonzalez2025two}.

\section*{Discussion}
We reported a quantum dot based photon-emitter interface in the original telecom band (O-band), exhibiting Fourier-limited optical linewidths, low blinking, high count rate and high indistinguishability. This is the first time that a telecom photon-emitter interface is developed with performance specs en route to the demanding requirements of fault tolerant photonic quantum computing. 

The results presented in this work lay the foundation of a new generation of photonic quantum systems ranging from deterministic photon-photon nonlinear devices to spin-photon interfaces and cluster-state generation, now made compatible with low-loss optical fibers and silicon photonics. The potential is significant: low-loss fibers ($\sim0.32$ dB/km) enable storing more than a thousand O-band photons with $<2$~\% loss (for example, a 300-m-long fiber offers around \SI{1.5}{\micro\s} delay time, which roughly corresponds to 1500 photons at 1~GHz repetition rate), significantly extending the operation time for feed-forward operations or loop-based boson sampling architectures \cite{aghaee2025scaling}. Furthermore, employing heterogeneous integration techniques, such as micro-transfer printing or wafer bonding \cite{zhang2019iii,davanco2017heterogeneous,salamon2025electrical}, QDs can be directly integrated into silicon-on-insulator photonic integrated circuits, offering a mature, CMOS-compatible, and industry-ready technology that benefits from the advantage of well-established commercial foundries. 
Another important direction is that O-band operation also enables direct deployment in passive optical network and data centers \cite{bernal202412} without resorting to bulky and expensive frequency conversion setups, offering near-zero dispersion in short- and intermediate-range urban quantum networks where the O-band enjoys little signal contamination from adjacent datacom fibers.
Finally, the availability of emitters at the telecom wavelengths opens a whole new paradigm of opportunities for the development of GaAs-based quantum photonic integrated circuits, with losses approaching 1~dB/cm \cite{ottaviano2016,chang2020}, compatibility with superconducting single-photon detectors \cite{sprengers2011waveguide}, and a native linear electro-optic coefficient, together enabling light generation, high-speed modulation, and detection in a single circuit platform \cite{uppu2021Quantum‑dot‑based}.

\section*{Online Methods}
\subsection*{Growth of the quantum dot samples}
The device heterostructure was grown by molecular beam epitaxy (MBE) on a (100) GaAs 3’’ wafer with the QDs embedded in a 224-nm-thick GaAs membrane with a doped $p$-$i$-$n$ junction on top of a 865-nm-thick Al$_{0.76}$Ga$_{0.24}$As sacrificial layer and a distributed Bragg reflector (DBR) composed of AlAs/GaAs targeting a center wavelength of \SI{1.3}{\micro\meter}. 
The detailed layer stack is provided in Supplementary Table~S1.

The indium arsenide (InAs) QDs were overgrown with a \SI{7}{\nano\meter}  $\mathrm{In}_{0.3}\mathrm{Ga}_{0.7}\mathrm{As}$ strain reduction layer (SRL). The SRL effectively shifts the QD emission wavelength to \SI{1.3}{\micro\meter} \cite{NISHI1999}. Low charge noise is achieved through charge stabilization in a heterostructure that is designed without artificial charge traps \cite{LUDWIG2017} and a strong focus on material quality. GaAs quality is monitored via periodic mobility measurements in dedicated high-electron mobility transistor (HEMT) heterostructure samples. The AlAs quality is checked with photoluminescence measurements on GaAs/AlGaAs quantum well (QW) samples \cite{NGUYEN2020}. Specifically, we observed that purified aluminium leads to smooth interfaces, which is particularly important for the DBR's effectiveness, and to minimize deep trap states in the AlGaAs barrier immediately above the QDs. 

\subsection*{Growth quality control}
During the growth optimization, quality-control samples were grown and characterized to monitor material properties. The optical and structural properties of GaAs and AlGaAs layers were investigated by the growth of quantum wells (QWs) with varying thicknesses.
Photoluminescence (PL) brightness from each QW sample was measured and normalized to a stable high-performance reference sample \cite{NGUYEN2020}.
An important finding during this study was the direct correlation between high QW emission brightness and low surface roughness (Fig.~S1). 
Smooth interfaces are known to improve the optical quality of QWs \cite{CHIN1998} and avoid dislocations. We therefore optimized our growth conditions to minimize interface roughness during growth.

We performed scanning transmission electron microscopy (STEM) images of the sample used in this work, focusing on the AlAs/GaAs interface in the DBR region. Fig.~S2 clearly demonstrates sharp and well-defined interfaces consistent with a smooth surface with a root-mean-square (RMS) roughness of \SI{224}{\pico\meter} as extracted by atomic force microscopy (AFM) measured with a Park System NX20 in non-contact mode, which confirms the high quality of the material growth.

\subsection*{Optimization of the aluminum quality}
A critical aspect of the growth process, that affects the entire sample quality, is the quality of the aluminum cell during epitaxial growth, where the risk of cross-contamination leads to higher roughness and consequently lower QD/QW brightness. The standard approach is to overheat the cells over several hours, which however leads to cross-contamination with the gallium cell. Our MBE system utilizes ion getter pumps, which have a comparatively lower pumping speed than typical cryopumps and may not efficiently evacuate volatile contaminants evaporated during the Al cell overheat. To counteract the cross contamination, we operate a titanium sublimation pump during the overheat without Ga at a frequency of \SI{0.5}{\min^{-1}}, and Al overheat temperature was set to \SI{135}{\kelvin} above growth temperature. After 3 iterations, a relative brightness of \SI{56}{\percent} with an RMS surface roughness of \SI{238}{\pico\meter} was achieved. Finally, the arsenic cell is purified by overheating it at \SI{900}{\degreeCelsius} for two full days, which enables reaching a relative intensity of \SI{103}{\percent} and a RMS of \SI{188}{\pico\meter}, i.e., the brightness is even stronger than the reference sample. Once these conditions were met, the QD sample was grown according to the layer stack of Supplementary Table~S1.

\subsection*{Device design and fabrication}
The membrane thickness is simply scaled from the standard 160~nm employed at 930~nm \cite{uppu2020scalable} to 224~nm at 1300~nm by maintaining the ratio and the photonic components including the shallow-etched grating coupler (50~nm etch depth), photonic crystals (lattice spacing $a=338$~nm and $r=0.27a$) and tethers are designed using the refractive index of GaAs at cryogenic temperature ($n_\text{GaAs,1300} = 3.35$).
The shallow-etched focusing grating couplers are optimized following the procedure outlined in ref. \cite{zhou2018high} for a central wavelength of \SI{1.3}{\micro\m}. Three-dimensional finite-difference time-domain simulations using the open-source software MEEP \cite{oskooi2010meep} of the entire gratings were carried out to optimize the distance between the grating and the bottom Bragg mirror, resulting in an 865~nm sacrificial layer thickness.

The electrical contacts are fabricated by physical vapor deposition in an electron-beam evaporator. We employ a Cr/Au bi-layer stack for $p$-type contacts and we deposit Ni/Ge/Au and perform rapid thermal annealing for $n$-type contacts. 
Shallow-etched focusing gratings and isolation trenches through the $p$-layer are patterned by electron-beam lithography and etched via reactive ion etching (RIE) in a BCl$_3$/Ar chemistry. In a separate step, waveguides and photonic crystals are deeply etched using inductively coupled plasma (ICP) reactive-ion etching (RIE) in a BCl$_3$/Cl$_2$/Ar chemistry, and the sacrificial AlGaAs layer is etched in hydrofluoric acid and finally dried using critical-point drying. A stitched microscope image of the entire chip is shown in Supplementary Fig.~S3. The chip is wire-bonded to a printed circuit board and mounted in the cryostat with coaxial electrical feedthroughs.

\subsection*{Resonance fluorescence measurements}
The sample is characterized in a closed-cycle cryostat (AttoDRY 800) at 4.3~K. We employ resonance fluorescence (RF) with a tunable continuous wave O-band laser (Santec TSL-570) to drive and characterize the emitters. To separate the excitation beam from the collection beam we cross-polarize the excitation spot on the waveguide from the collection grating, achieving both spatial and polarization extinction.
The RF map in Fig.~\ref{fig:2} is obtained by stepping the excitation laser wavelength from \SI{1295}{nm} to \SI{1300}{nm} in \SI{10}{pm} steps. At each step the gate voltage is stepped from $-\SI{450}{mV}$ to $-\SI{290}{mV}$ in discrete \SI{50}{\micro\V} steps with 20~ms settling time and 100~ms integration time, i.e., the data is recorded across 54 hours without changing the alignment.

Our pulsed laser (PicusQ, Refined Lasers) is an integrated fiber optical parametric oscillator, which emits a signal (780--980~nm) and idler (1100--1550~nm) pulse from a $\sim1046$~nm pump, using the idler to lock the target signal wavelength. This means that the laser wavelength drifts on the time-scale of tens of seconds. For resonant excitation, we set the signal around $\lambda_s\sim876.0$~nm, adjusting it to keep the idler stable. The idler pulses are already $\sim\SI{10}{ps}$ un-stretched, and we use only a minor filtering with a 40~GHz stretcher to avoid too long pulses, which would cause re-excitation. Using faster (spectrally broader) initial pulses or locking the idler directly would improve the pump laser stability.
In fact, the high count rate of our emitter allows direct aligning to the $g^{(2)}(0)$ leading to the result shown in the main text, however, the drift in the pulsed laser during alignment of the HOM setup caused the $g^{(2)}(0)$ to increase to $0.0773(3)$ when recorded by blocking one of the interferometer arms immediately after the HOM measurement presented in the main text.


\begin{thebibliography}{55}%
\makeatletter
\providecommand \@ifxundefined [1]{%
 \@ifx{#1\undefined}
}%
\providecommand \@ifnum [1]{%
 \ifnum #1\expandafter \@firstoftwo
 \else \expandafter \@secondoftwo
 \fi
}%
\providecommand \@ifx [1]{%
 \ifx #1\expandafter \@firstoftwo
 \else \expandafter \@secondoftwo
 \fi
}%
\providecommand \natexlab [1]{#1}%
\providecommand \enquote  [1]{``#1''}%
\providecommand \bibnamefont  [1]{#1}%
\providecommand \bibfnamefont [1]{#1}%
\providecommand \citenamefont [1]{#1}%
\providecommand \href@noop [0]{\@secondoftwo}%
\providecommand \href [0]{\begingroup \@sanitize@url \@href}%
\providecommand \@href[1]{\@@startlink{#1}\@@href}%
\providecommand \@@href[1]{\endgroup#1\@@endlink}%
\providecommand \@sanitize@url [0]{\catcode `\\12\catcode `\$12\catcode `\&12\catcode `\#12\catcode `\^12\catcode `\_12\catcode `\%12\relax}%
\providecommand \@@startlink[1]{}%
\providecommand \@@endlink[0]{}%
\providecommand \url  [0]{\begingroup\@sanitize@url \@url }%
\providecommand \@url [1]{\endgroup\@href {#1}{\urlprefix }}%
\providecommand \urlprefix  [0]{URL }%
\providecommand \Eprint [0]{\href }%
\providecommand \doibase [0]{https://doi.org/}%
\providecommand \selectlanguage [0]{\@gobble}%
\providecommand \bibinfo  [0]{\@secondoftwo}%
\providecommand \bibfield  [0]{\@secondoftwo}%
\providecommand \translation [1]{[#1]}%
\providecommand \BibitemOpen [0]{}%
\providecommand \bibitemStop [0]{}%
\providecommand \bibitemNoStop [0]{.\EOS\space}%
\providecommand \EOS [0]{\spacefactor3000\relax}%
\providecommand \BibitemShut  [1]{\csname bibitem#1\endcsname}%
\let\auto@bib@innerbib\@empty
\bibitem [{\citenamefont {Wang}\ \emph {et~al.}(2025)\citenamefont {Wang}, \citenamefont {Ralph}, \citenamefont {Renema}, \citenamefont {Lu},\ and\ \citenamefont {Pan}}]{wang2025scalable}%
  \BibitemOpen
  \bibfield  {author} {\bibinfo {author} {\bibfnamefont {H.}~\bibnamefont {Wang}}, \bibinfo {author} {\bibfnamefont {T.~C.}\ \bibnamefont {Ralph}}, \bibinfo {author} {\bibfnamefont {J.~J.}\ \bibnamefont {Renema}}, \bibinfo {author} {\bibfnamefont {C.-Y.}\ \bibnamefont {Lu}},\ and\ \bibinfo {author} {\bibfnamefont {J.-W.}\ \bibnamefont {Pan}},\ }\bibfield  {title} {\bibinfo {title} {Scalable photonic quantum technologies},\ }\href@noop {} {\bibfield  {journal} {\bibinfo  {journal} {Nat. Mater.}\ } (\bibinfo {year} {2025})}\BibitemShut {NoStop}%
\bibitem [{\citenamefont {Yin}\ \emph {et~al.}(2017)\citenamefont {Yin}, \citenamefont {Cao}, \citenamefont {Li}, \citenamefont {Liao}, \citenamefont {Zhang}, \citenamefont {Ren}, \citenamefont {Cai}, \citenamefont {Liu}, \citenamefont {Li}, \citenamefont {Dai} \emph {et~al.}}]{yin2017satellite}%
  \BibitemOpen
  \bibfield  {author} {\bibinfo {author} {\bibfnamefont {J.}~\bibnamefont {Yin}}, \bibinfo {author} {\bibfnamefont {Y.}~\bibnamefont {Cao}}, \bibinfo {author} {\bibfnamefont {Y.-H.}\ \bibnamefont {Li}}, \bibinfo {author} {\bibfnamefont {S.-K.}\ \bibnamefont {Liao}}, \bibinfo {author} {\bibfnamefont {L.}~\bibnamefont {Zhang}}, \bibinfo {author} {\bibfnamefont {J.-G.}\ \bibnamefont {Ren}}, \bibinfo {author} {\bibfnamefont {W.-Q.}\ \bibnamefont {Cai}}, \bibinfo {author} {\bibfnamefont {W.-Y.}\ \bibnamefont {Liu}}, \bibinfo {author} {\bibfnamefont {B.}~\bibnamefont {Li}}, \bibinfo {author} {\bibfnamefont {H.}~\bibnamefont {Dai}}, \emph {et~al.},\ }\bibfield  {title} {\bibinfo {title} {Satellite-based entanglement distribution over 1200 kilometers},\ }\href@noop {} {\bibfield  {journal} {\bibinfo  {journal} {Science}\ }\textbf {\bibinfo {volume} {356}},\ \bibinfo {pages} {1140} (\bibinfo {year} {2017})}\BibitemShut {NoStop}%
\bibitem [{\citenamefont {Wang}\ \emph {et~al.}(2020)\citenamefont {Wang}, \citenamefont {Sciarrino}, \citenamefont {Laing},\ and\ \citenamefont {Thompson}}]{wang2020integrated}%
  \BibitemOpen
  \bibfield  {author} {\bibinfo {author} {\bibfnamefont {J.}~\bibnamefont {Wang}}, \bibinfo {author} {\bibfnamefont {F.}~\bibnamefont {Sciarrino}}, \bibinfo {author} {\bibfnamefont {A.}~\bibnamefont {Laing}},\ and\ \bibinfo {author} {\bibfnamefont {M.~G.}\ \bibnamefont {Thompson}},\ }\bibfield  {title} {\bibinfo {title} {Integrated photonic quantum technologies},\ }\href@noop {} {\bibfield  {journal} {\bibinfo  {journal} {Nat. Photon.}\ }\textbf {\bibinfo {volume} {14}},\ \bibinfo {pages} {273} (\bibinfo {year} {2020})}\BibitemShut {NoStop}%
\bibitem [{\citenamefont {Lu}\ and\ \citenamefont {Pan}(2021)}]{lu2021quantum}%
  \BibitemOpen
  \bibfield  {author} {\bibinfo {author} {\bibfnamefont {C.-Y.}\ \bibnamefont {Lu}}\ and\ \bibinfo {author} {\bibfnamefont {J.-W.}\ \bibnamefont {Pan}},\ }\bibfield  {title} {\bibinfo {title} {Quantum-dot single-photon sources for the quantum internet},\ }\href@noop {} {\bibfield  {journal} {\bibinfo  {journal} {Nat. Nanotechnol.}\ }\textbf {\bibinfo {volume} {16}},\ \bibinfo {pages} {1294} (\bibinfo {year} {2021})}\BibitemShut {NoStop}%
\bibitem [{\citenamefont {Uppu}\ \emph {et~al.}(2021)\citenamefont {Uppu}, \citenamefont {Midolo}, \citenamefont {Zhou}, \citenamefont {Carolan},\ and\ \citenamefont {Lodahl}}]{uppu2021Quantum‑dot‑based}%
  \BibitemOpen
  \bibfield  {author} {\bibinfo {author} {\bibfnamefont {R.}~\bibnamefont {Uppu}}, \bibinfo {author} {\bibfnamefont {L.}~\bibnamefont {Midolo}}, \bibinfo {author} {\bibfnamefont {X.}~\bibnamefont {Zhou}}, \bibinfo {author} {\bibfnamefont {J.}~\bibnamefont {Carolan}},\ and\ \bibinfo {author} {\bibfnamefont {P.}~\bibnamefont {Lodahl}},\ }\bibfield  {title} {\bibinfo {title} {Quantum‑dot‑based deterministic photon‑emitter interfaces for scalable photonic quantum technology},\ }\href@noop {} {\bibfield  {journal} {\bibinfo  {journal} {Nat. Nanotechnol.}\ }\textbf {\bibinfo {volume} {16}},\ \bibinfo {pages} {1308} (\bibinfo {year} {2021})}\BibitemShut {NoStop}%
\bibitem [{\citenamefont {Azuma}\ \emph {et~al.}(2015)\citenamefont {Azuma}, \citenamefont {Tamaki},\ and\ \citenamefont {Lo}}]{azuma2015all}%
  \BibitemOpen
  \bibfield  {author} {\bibinfo {author} {\bibfnamefont {K.}~\bibnamefont {Azuma}}, \bibinfo {author} {\bibfnamefont {K.}~\bibnamefont {Tamaki}},\ and\ \bibinfo {author} {\bibfnamefont {H.-K.}\ \bibnamefont {Lo}},\ }\bibfield  {title} {\bibinfo {title} {All-photonic quantum repeaters},\ }\href@noop {} {\bibfield  {journal} {\bibinfo  {journal} {Nat. Commun.}\ }\textbf {\bibinfo {volume} {6}},\ \bibinfo {pages} {6787} (\bibinfo {year} {2015})}\BibitemShut {NoStop}%
\bibitem [{\citenamefont {Borregaard}\ \emph {et~al.}(2020)\citenamefont {Borregaard}, \citenamefont {Pichler}, \citenamefont {Schr{\"o}der}, \citenamefont {Lukin}, \citenamefont {Lodahl},\ and\ \citenamefont {S{\o}rensen}}]{borregaard2020one}%
  \BibitemOpen
  \bibfield  {author} {\bibinfo {author} {\bibfnamefont {J.}~\bibnamefont {Borregaard}}, \bibinfo {author} {\bibfnamefont {H.}~\bibnamefont {Pichler}}, \bibinfo {author} {\bibfnamefont {T.}~\bibnamefont {Schr{\"o}der}}, \bibinfo {author} {\bibfnamefont {M.~D.}\ \bibnamefont {Lukin}}, \bibinfo {author} {\bibfnamefont {P.}~\bibnamefont {Lodahl}},\ and\ \bibinfo {author} {\bibfnamefont {A.~S.}\ \bibnamefont {S{\o}rensen}},\ }\bibfield  {title} {\bibinfo {title} {One-way quantum repeater based on near-deterministic photon-emitter interfaces},\ }\href@noop {} {\bibfield  {journal} {\bibinfo  {journal} {Phys. Rev. X}\ }\textbf {\bibinfo {volume} {10}},\ \bibinfo {pages} {021071} (\bibinfo {year} {2020})}\BibitemShut {NoStop}%
\bibitem [{\citenamefont {Ko{\l}ody{\'n}ski}\ \emph {et~al.}(2020)\citenamefont {Ko{\l}ody{\'n}ski}, \citenamefont {M{\'a}ttar}, \citenamefont {Skrzypczyk}, \citenamefont {Woodhead}, \citenamefont {Cavalcanti}, \citenamefont {Banaszek},\ and\ \citenamefont {Ac{\'\i}n}}]{kolodynski2020device}%
  \BibitemOpen
  \bibfield  {author} {\bibinfo {author} {\bibfnamefont {J.}~\bibnamefont {Ko{\l}ody{\'n}ski}}, \bibinfo {author} {\bibfnamefont {A.}~\bibnamefont {M{\'a}ttar}}, \bibinfo {author} {\bibfnamefont {P.}~\bibnamefont {Skrzypczyk}}, \bibinfo {author} {\bibfnamefont {E.}~\bibnamefont {Woodhead}}, \bibinfo {author} {\bibfnamefont {D.}~\bibnamefont {Cavalcanti}}, \bibinfo {author} {\bibfnamefont {K.}~\bibnamefont {Banaszek}},\ and\ \bibinfo {author} {\bibfnamefont {A.}~\bibnamefont {Ac{\'\i}n}},\ }\bibfield  {title} {\bibinfo {title} {Device-independent quantum key distribution with single-photon sources},\ }\href@noop {} {\bibfield  {journal} {\bibinfo  {journal} {Quantum}\ }\textbf {\bibinfo {volume} {4}},\ \bibinfo {pages} {260} (\bibinfo {year} {2020})}\BibitemShut {NoStop}%
\bibitem [{\citenamefont {Alexander}\ \emph {et~al.}(2025)\citenamefont {Alexander}, \citenamefont {Benyamini}, \citenamefont {Black}, \citenamefont {Bonneau}, \citenamefont {Burgos}, \citenamefont {Burridge}, \citenamefont {Cable}, \citenamefont {Campbell}, \citenamefont {Catalano}, \citenamefont {Ceballos} \emph {et~al.}}]{psiquantum2025manufacturable}%
  \BibitemOpen
  \bibfield  {author} {\bibinfo {author} {\bibfnamefont {K.}~\bibnamefont {Alexander}}, \bibinfo {author} {\bibfnamefont {A.}~\bibnamefont {Benyamini}}, \bibinfo {author} {\bibfnamefont {D.}~\bibnamefont {Black}}, \bibinfo {author} {\bibfnamefont {D.}~\bibnamefont {Bonneau}}, \bibinfo {author} {\bibfnamefont {S.}~\bibnamefont {Burgos}}, \bibinfo {author} {\bibfnamefont {B.}~\bibnamefont {Burridge}}, \bibinfo {author} {\bibfnamefont {H.}~\bibnamefont {Cable}}, \bibinfo {author} {\bibfnamefont {G.}~\bibnamefont {Campbell}}, \bibinfo {author} {\bibfnamefont {G.}~\bibnamefont {Catalano}}, \bibinfo {author} {\bibfnamefont {A.}~\bibnamefont {Ceballos}}, \emph {et~al.},\ }\bibfield  {title} {\bibinfo {title} {A manufacturable platform for photonic quantum computing},\ }\href@noop {} {\bibfield  {journal} {\bibinfo  {journal} {Nature}\ }\textbf {\bibinfo {volume} {641}},\ \bibinfo {pages} {876} (\bibinfo {year} {2025})}\BibitemShut {NoStop}%
\bibitem [{\citenamefont {Chan}\ \emph {et~al.}(2025)\citenamefont {Chan}, \citenamefont {Capatos}, \citenamefont {Lodahl}, \citenamefont {S{\o}rensen},\ and\ \citenamefont {Paesani}}]{chan2025practical}%
  \BibitemOpen
  \bibfield  {author} {\bibinfo {author} {\bibfnamefont {M.~L.}\ \bibnamefont {Chan}}, \bibinfo {author} {\bibfnamefont {A.~A.}\ \bibnamefont {Capatos}}, \bibinfo {author} {\bibfnamefont {P.}~\bibnamefont {Lodahl}}, \bibinfo {author} {\bibfnamefont {A.~S.}\ \bibnamefont {S{\o}rensen}},\ and\ \bibinfo {author} {\bibfnamefont {S.}~\bibnamefont {Paesani}},\ }\bibfield  {title} {\bibinfo {title} {Practical blueprint for low-depth photonic quantum computing with quantum dots},\ }\href@noop {} {\bibfield  {journal} {\bibinfo  {journal} {Preprint at arXiv:2507.16152}\ } (\bibinfo {year} {2025})}\BibitemShut {NoStop}%
\bibitem [{\citenamefont {De~Leon}\ \emph {et~al.}(2021)\citenamefont {De~Leon}, \citenamefont {Itoh}, \citenamefont {Kim}, \citenamefont {Mehta}, \citenamefont {Northup}, \citenamefont {Paik}, \citenamefont {Palmer}, \citenamefont {Samarth}, \citenamefont {Sangtawesin},\ and\ \citenamefont {Steuerman}}]{de2021materials}%
  \BibitemOpen
  \bibfield  {author} {\bibinfo {author} {\bibfnamefont {N.~P.}\ \bibnamefont {De~Leon}}, \bibinfo {author} {\bibfnamefont {K.~M.}\ \bibnamefont {Itoh}}, \bibinfo {author} {\bibfnamefont {D.}~\bibnamefont {Kim}}, \bibinfo {author} {\bibfnamefont {K.~K.}\ \bibnamefont {Mehta}}, \bibinfo {author} {\bibfnamefont {T.~E.}\ \bibnamefont {Northup}}, \bibinfo {author} {\bibfnamefont {H.}~\bibnamefont {Paik}}, \bibinfo {author} {\bibfnamefont {B.}~\bibnamefont {Palmer}}, \bibinfo {author} {\bibfnamefont {N.}~\bibnamefont {Samarth}}, \bibinfo {author} {\bibfnamefont {S.}~\bibnamefont {Sangtawesin}},\ and\ \bibinfo {author} {\bibfnamefont {D.~W.}\ \bibnamefont {Steuerman}},\ }\bibfield  {title} {\bibinfo {title} {Materials challenges and opportunities for quantum computing hardware},\ }\href@noop {} {\bibfield  {journal} {\bibinfo  {journal} {Science}\ }\textbf {\bibinfo {volume} {372}},\ \bibinfo {pages} {eabb2823} (\bibinfo {year} {2021})}\BibitemShut {NoStop}%
\bibitem [{\citenamefont {Lodahl}\ \emph {et~al.}(2015)\citenamefont {Lodahl}, \citenamefont {Mahmoodian},\ and\ \citenamefont {Stobbe}}]{lodahl2015Interfacing}%
  \BibitemOpen
  \bibfield  {author} {\bibinfo {author} {\bibfnamefont {P.}~\bibnamefont {Lodahl}}, \bibinfo {author} {\bibfnamefont {S.}~\bibnamefont {Mahmoodian}},\ and\ \bibinfo {author} {\bibfnamefont {S.}~\bibnamefont {Stobbe}},\ }\bibfield  {title} {\bibinfo {title} {Interfacing single photons and single quantum dots with photonic nanostructures},\ }\href@noop {} {\bibfield  {journal} {\bibinfo  {journal} {Rev. Mod. Phys.}\ }\textbf {\bibinfo {volume} {87}},\ \bibinfo {pages} {347} (\bibinfo {year} {2015})}\BibitemShut {NoStop}%
\bibitem [{\citenamefont {Uppu}\ \emph {et~al.}(2020)\citenamefont {Uppu}, \citenamefont {Midolo}, \citenamefont {Zhou}, \citenamefont {Carolan},\ and\ \citenamefont {Lodahl}}]{uppu2020scalable}%
  \BibitemOpen
  \bibfield  {author} {\bibinfo {author} {\bibfnamefont {R.}~\bibnamefont {Uppu}}, \bibinfo {author} {\bibfnamefont {L.}~\bibnamefont {Midolo}}, \bibinfo {author} {\bibfnamefont {X.}~\bibnamefont {Zhou}}, \bibinfo {author} {\bibfnamefont {J.}~\bibnamefont {Carolan}},\ and\ \bibinfo {author} {\bibfnamefont {P.}~\bibnamefont {Lodahl}},\ }\bibfield  {title} {\bibinfo {title} {Scalable integrated single-photon source},\ }\href@noop {} {\bibfield  {journal} {\bibinfo  {journal} {Sci. Adv.}\ }\textbf {\bibinfo {volume} {6}},\ \bibinfo {pages} {eabc8268} (\bibinfo {year} {2020})}\BibitemShut {NoStop}%
\bibitem [{\citenamefont {Tomm}\ \emph {et~al.}(2021)\citenamefont {Tomm}, \citenamefont {Javadi}, \citenamefont {Antoniadis}, \citenamefont {Najer}, \citenamefont {L{\"o}bl}, \citenamefont {Korsch}, \citenamefont {Schott}, \citenamefont {Valentin}, \citenamefont {Wieck}, \citenamefont {Ludwig},\ and\ \citenamefont {Warburton}}]{tomm2021bright}%
  \BibitemOpen
  \bibfield  {author} {\bibinfo {author} {\bibfnamefont {N.}~\bibnamefont {Tomm}}, \bibinfo {author} {\bibfnamefont {A.}~\bibnamefont {Javadi}}, \bibinfo {author} {\bibfnamefont {N.~O.}\ \bibnamefont {Antoniadis}}, \bibinfo {author} {\bibfnamefont {D.}~\bibnamefont {Najer}}, \bibinfo {author} {\bibfnamefont {M.~C.}\ \bibnamefont {L{\"o}bl}}, \bibinfo {author} {\bibfnamefont {A.~R.}\ \bibnamefont {Korsch}}, \bibinfo {author} {\bibfnamefont {R.}~\bibnamefont {Schott}}, \bibinfo {author} {\bibfnamefont {S.~R.}\ \bibnamefont {Valentin}}, \bibinfo {author} {\bibfnamefont {A.~D.}\ \bibnamefont {Wieck}}, \bibinfo {author} {\bibfnamefont {A.}~\bibnamefont {Ludwig}},\ and\ \bibinfo {author} {\bibfnamefont {R.~J.}\ \bibnamefont {Warburton}},\ }\bibfield  {title} {\bibinfo {title} {A bright and fast source of coherent single photons},\ }\href@noop {} {\bibfield  {journal} {\bibinfo  {journal} {Nat. Nanotechnol.}\ }\textbf {\bibinfo {volume} {16}},\ \bibinfo {pages} {399} (\bibinfo {year} {2021})}\BibitemShut {NoStop}%
\bibitem [{\citenamefont {Ding}\ \emph {et~al.}(2025)\citenamefont {Ding}, \citenamefont {Guo}, \citenamefont {Xu}, \citenamefont {Liu}, \citenamefont {Zou}, \citenamefont {Zhao}, \citenamefont {Ge}, \citenamefont {Zhang}, \citenamefont {Liu}, \citenamefont {Wang}, \citenamefont {Chen}, \citenamefont {Wang}, \citenamefont {He}, \citenamefont {Huo}, \citenamefont {Lu},\ and\ \citenamefont {Pan}}]{ding2025high}%
  \BibitemOpen
  \bibfield  {author} {\bibinfo {author} {\bibfnamefont {X.}~\bibnamefont {Ding}}, \bibinfo {author} {\bibfnamefont {Y.-P.}\ \bibnamefont {Guo}}, \bibinfo {author} {\bibfnamefont {M.-C.}\ \bibnamefont {Xu}}, \bibinfo {author} {\bibfnamefont {R.-Z.}\ \bibnamefont {Liu}}, \bibinfo {author} {\bibfnamefont {G.-Y.}\ \bibnamefont {Zou}}, \bibinfo {author} {\bibfnamefont {J.-Y.}\ \bibnamefont {Zhao}}, \bibinfo {author} {\bibfnamefont {Z.-X.}\ \bibnamefont {Ge}}, \bibinfo {author} {\bibfnamefont {Q.-H.}\ \bibnamefont {Zhang}}, \bibinfo {author} {\bibfnamefont {H.-L.}\ \bibnamefont {Liu}}, \bibinfo {author} {\bibfnamefont {L.-J.}\ \bibnamefont {Wang}}, \bibinfo {author} {\bibfnamefont {M.-C.}\ \bibnamefont {Chen}}, \bibinfo {author} {\bibfnamefont {H.}~\bibnamefont {Wang}}, \bibinfo {author} {\bibfnamefont {Y.-M.}\ \bibnamefont {He}}, \bibinfo {author} {\bibfnamefont {Y.-H.}\ \bibnamefont {Huo}}, \bibinfo {author} {\bibfnamefont {C.-Y.}\ \bibnamefont {Lu}},\ and\ \bibinfo {author} {\bibfnamefont {J.-W.}\ \bibnamefont
  {Pan}},\ }\bibfield  {title} {\bibinfo {title} {High-efficiency single-photon source above the loss-tolerant threshold for efficient linear optical quantum computing},\ }\href@noop {} {\bibfield  {journal} {\bibinfo  {journal} {Nat. Photon.}\ }\textbf {\bibinfo {volume} {19}},\ \bibinfo {pages} {387} (\bibinfo {year} {2025})}\BibitemShut {NoStop}%
\bibitem [{\citenamefont {Alloing}\ \emph {et~al.}(2005)\citenamefont {Alloing}, \citenamefont {Zinoni}, \citenamefont {Zwiller}, \citenamefont {Li}, \citenamefont {Monat}, \citenamefont {Gobet}, \citenamefont {Buchs}, \citenamefont {Fiore}, \citenamefont {Pelucchi},\ and\ \citenamefont {Kapon}}]{alloing2005growth}%
  \BibitemOpen
  \bibfield  {author} {\bibinfo {author} {\bibfnamefont {B.}~\bibnamefont {Alloing}}, \bibinfo {author} {\bibfnamefont {C.}~\bibnamefont {Zinoni}}, \bibinfo {author} {\bibfnamefont {V.}~\bibnamefont {Zwiller}}, \bibinfo {author} {\bibfnamefont {L.}~\bibnamefont {Li}}, \bibinfo {author} {\bibfnamefont {C.}~\bibnamefont {Monat}}, \bibinfo {author} {\bibfnamefont {M.}~\bibnamefont {Gobet}}, \bibinfo {author} {\bibfnamefont {G.}~\bibnamefont {Buchs}}, \bibinfo {author} {\bibfnamefont {A.}~\bibnamefont {Fiore}}, \bibinfo {author} {\bibfnamefont {E.}~\bibnamefont {Pelucchi}},\ and\ \bibinfo {author} {\bibfnamefont {E.}~\bibnamefont {Kapon}},\ }\bibfield  {title} {\bibinfo {title} {Growth and characterization of single quantum dots emitting at 1300 nm},\ }\href@noop {} {\bibfield  {journal} {\bibinfo  {journal} {Appl. Phys. Lett.}\ }\textbf {\bibinfo {volume} {86}} (\bibinfo {year} {2005})}\BibitemShut {NoStop}%
\bibitem [{\citenamefont {Holewa}\ \emph {et~al.}(2025)\citenamefont {Holewa}, \citenamefont {Reiserer}, \citenamefont {Heindel}, \citenamefont {Sanguinetti}, \citenamefont {Huck},\ and\ \citenamefont {Semenova}}]{holewa2025solid}%
  \BibitemOpen
  \bibfield  {author} {\bibinfo {author} {\bibfnamefont {P.}~\bibnamefont {Holewa}}, \bibinfo {author} {\bibfnamefont {A.}~\bibnamefont {Reiserer}}, \bibinfo {author} {\bibfnamefont {T.}~\bibnamefont {Heindel}}, \bibinfo {author} {\bibfnamefont {S.}~\bibnamefont {Sanguinetti}}, \bibinfo {author} {\bibfnamefont {A.}~\bibnamefont {Huck}},\ and\ \bibinfo {author} {\bibfnamefont {E.}~\bibnamefont {Semenova}},\ }\bibfield  {title} {\bibinfo {title} {Solid-state single-photon sources operating in the telecom wavelength range},\ }\href@noop {} {\bibfield  {journal} {\bibinfo  {journal} {Nanophotonics}\ }\textbf {\bibinfo {volume} {14}},\ \bibinfo {pages} {1729} (\bibinfo {year} {2025})}\BibitemShut {NoStop}%
\bibitem [{\citenamefont {Nawrath}\ \emph {et~al.}(2023)\citenamefont {Nawrath}, \citenamefont {Joos}, \citenamefont {Kolatschek}, \citenamefont {Bauer}, \citenamefont {Pruy}, \citenamefont {Hornung}, \citenamefont {Fischer}, \citenamefont {Huang}, \citenamefont {Vijayan}, \citenamefont {Sittig} \emph {et~al.}}]{nawrath2023bright}%
  \BibitemOpen
  \bibfield  {author} {\bibinfo {author} {\bibfnamefont {C.}~\bibnamefont {Nawrath}}, \bibinfo {author} {\bibfnamefont {R.}~\bibnamefont {Joos}}, \bibinfo {author} {\bibfnamefont {S.}~\bibnamefont {Kolatschek}}, \bibinfo {author} {\bibfnamefont {S.}~\bibnamefont {Bauer}}, \bibinfo {author} {\bibfnamefont {P.}~\bibnamefont {Pruy}}, \bibinfo {author} {\bibfnamefont {F.}~\bibnamefont {Hornung}}, \bibinfo {author} {\bibfnamefont {J.}~\bibnamefont {Fischer}}, \bibinfo {author} {\bibfnamefont {J.}~\bibnamefont {Huang}}, \bibinfo {author} {\bibfnamefont {P.}~\bibnamefont {Vijayan}}, \bibinfo {author} {\bibfnamefont {R.}~\bibnamefont {Sittig}}, \emph {et~al.},\ }\bibfield  {title} {\bibinfo {title} {Bright source of purcell-enhanced, triggered, single photons in the telecom c-band},\ }\href@noop {} {\bibfield  {journal} {\bibinfo  {journal} {Adv. Quantum Technol.}\ }\textbf {\bibinfo {volume} {6}},\ \bibinfo {pages} {2300111} (\bibinfo {year} {2023})}\BibitemShut {NoStop}%
\bibitem [{\citenamefont {Joos}\ \emph {et~al.}(2024)\citenamefont {Joos}, \citenamefont {Bauer}, \citenamefont {Rupp}, \citenamefont {Kolatschek}, \citenamefont {Fischer}, \citenamefont {Nawrath}, \citenamefont {Vijayan}, \citenamefont {Sittig}, \citenamefont {Jetter}, \citenamefont {Portalupi} \emph {et~al.}}]{joos2024coherently}%
  \BibitemOpen
  \bibfield  {author} {\bibinfo {author} {\bibfnamefont {R.}~\bibnamefont {Joos}}, \bibinfo {author} {\bibfnamefont {S.}~\bibnamefont {Bauer}}, \bibinfo {author} {\bibfnamefont {C.}~\bibnamefont {Rupp}}, \bibinfo {author} {\bibfnamefont {S.}~\bibnamefont {Kolatschek}}, \bibinfo {author} {\bibfnamefont {W.}~\bibnamefont {Fischer}}, \bibinfo {author} {\bibfnamefont {C.}~\bibnamefont {Nawrath}}, \bibinfo {author} {\bibfnamefont {P.}~\bibnamefont {Vijayan}}, \bibinfo {author} {\bibfnamefont {R.}~\bibnamefont {Sittig}}, \bibinfo {author} {\bibfnamefont {M.}~\bibnamefont {Jetter}}, \bibinfo {author} {\bibfnamefont {S.~L.}\ \bibnamefont {Portalupi}}, \emph {et~al.},\ }\bibfield  {title} {\bibinfo {title} {Coherently and incoherently pumped telecom c-band single-photon source with high brightness and indistinguishability},\ }\href@noop {} {\bibfield  {journal} {\bibinfo  {journal} {Nano Lett.}\ }\textbf {\bibinfo {volume} {24}},\ \bibinfo {pages} {8626} (\bibinfo {year} {2024})}\BibitemShut {NoStop}%
\bibitem [{\citenamefont {Srocka}\ \emph {et~al.}(2020)\citenamefont {Srocka}, \citenamefont {Mrowi{\'n}ski}, \citenamefont {Gro{\ss}e}, \citenamefont {Von~Helversen}, \citenamefont {Heindel}, \citenamefont {Rodt},\ and\ \citenamefont {Reitzenstein}}]{srocka2020deterministically}%
  \BibitemOpen
  \bibfield  {author} {\bibinfo {author} {\bibfnamefont {N.}~\bibnamefont {Srocka}}, \bibinfo {author} {\bibfnamefont {P.}~\bibnamefont {Mrowi{\'n}ski}}, \bibinfo {author} {\bibfnamefont {J.}~\bibnamefont {Gro{\ss}e}}, \bibinfo {author} {\bibfnamefont {M.}~\bibnamefont {Von~Helversen}}, \bibinfo {author} {\bibfnamefont {T.}~\bibnamefont {Heindel}}, \bibinfo {author} {\bibfnamefont {S.}~\bibnamefont {Rodt}},\ and\ \bibinfo {author} {\bibfnamefont {S.}~\bibnamefont {Reitzenstein}},\ }\bibfield  {title} {\bibinfo {title} {Deterministically fabricated quantum dot single-photon source emitting indistinguishable photons in the telecom o-band},\ }\href@noop {} {\bibfield  {journal} {\bibinfo  {journal} {Appl. Phys. Lett.}\ }\textbf {\bibinfo {volume} {116}} (\bibinfo {year} {2020})}\BibitemShut {NoStop}%
\bibitem [{\citenamefont {Komza}\ \emph {et~al.}(2024)\citenamefont {Komza}, \citenamefont {Samutpraphoot}, \citenamefont {Odeh}, \citenamefont {Tang}, \citenamefont {Mathew}, \citenamefont {Chang}, \citenamefont {Song}, \citenamefont {Kim}, \citenamefont {Xiong}, \citenamefont {Hautier} \emph {et~al.}}]{komza2024indistinguishable}%
  \BibitemOpen
  \bibfield  {author} {\bibinfo {author} {\bibfnamefont {L.}~\bibnamefont {Komza}}, \bibinfo {author} {\bibfnamefont {P.}~\bibnamefont {Samutpraphoot}}, \bibinfo {author} {\bibfnamefont {M.}~\bibnamefont {Odeh}}, \bibinfo {author} {\bibfnamefont {Y.-L.}\ \bibnamefont {Tang}}, \bibinfo {author} {\bibfnamefont {M.}~\bibnamefont {Mathew}}, \bibinfo {author} {\bibfnamefont {J.}~\bibnamefont {Chang}}, \bibinfo {author} {\bibfnamefont {H.}~\bibnamefont {Song}}, \bibinfo {author} {\bibfnamefont {M.-K.}\ \bibnamefont {Kim}}, \bibinfo {author} {\bibfnamefont {Y.}~\bibnamefont {Xiong}}, \bibinfo {author} {\bibfnamefont {G.}~\bibnamefont {Hautier}}, \emph {et~al.},\ }\bibfield  {title} {\bibinfo {title} {Indistinguishable photons from an artificial atom in silicon photonics},\ }\href@noop {} {\bibfield  {journal} {\bibinfo  {journal} {Nat. Commun.}\ }\textbf {\bibinfo {volume} {15}},\ \bibinfo {pages} {6920} (\bibinfo {year} {2024})}\BibitemShut {NoStop}%
\bibitem [{\citenamefont {Simmons}(2024)}]{simmons2024scalable}%
  \BibitemOpen
  \bibfield  {author} {\bibinfo {author} {\bibfnamefont {S.}~\bibnamefont {Simmons}},\ }\bibfield  {title} {\bibinfo {title} {Scalable fault-tolerant quantum technologies with silicon color centers},\ }\href@noop {} {\bibfield  {journal} {\bibinfo  {journal} {PRX Quantum}\ }\textbf {\bibinfo {volume} {5}},\ \bibinfo {pages} {010102} (\bibinfo {year} {2024})}\BibitemShut {NoStop}%
\bibitem [{\citenamefont {Ourari}\ \emph {et~al.}(2023)\citenamefont {Ourari}, \citenamefont {Dusanowski}, \citenamefont {Horvath}, \citenamefont {Uysal}, \citenamefont {Phenicie}, \citenamefont {Stevenson}, \citenamefont {Raha}, \citenamefont {Chen}, \citenamefont {Cava}, \citenamefont {de~Leon} \emph {et~al.}}]{ourari2023indistinguishable}%
  \BibitemOpen
  \bibfield  {author} {\bibinfo {author} {\bibfnamefont {S.}~\bibnamefont {Ourari}}, \bibinfo {author} {\bibfnamefont {{\L}.}~\bibnamefont {Dusanowski}}, \bibinfo {author} {\bibfnamefont {S.~P.}\ \bibnamefont {Horvath}}, \bibinfo {author} {\bibfnamefont {M.~T.}\ \bibnamefont {Uysal}}, \bibinfo {author} {\bibfnamefont {C.~M.}\ \bibnamefont {Phenicie}}, \bibinfo {author} {\bibfnamefont {P.}~\bibnamefont {Stevenson}}, \bibinfo {author} {\bibfnamefont {M.}~\bibnamefont {Raha}}, \bibinfo {author} {\bibfnamefont {S.}~\bibnamefont {Chen}}, \bibinfo {author} {\bibfnamefont {R.~J.}\ \bibnamefont {Cava}}, \bibinfo {author} {\bibfnamefont {N.~P.}\ \bibnamefont {de~Leon}}, \emph {et~al.},\ }\bibfield  {title} {\bibinfo {title} {Indistinguishable telecom band photons from a single er ion in the solid state},\ }\href@noop {} {\bibfield  {journal} {\bibinfo  {journal} {Nature}\ }\textbf {\bibinfo {volume} {620}},\ \bibinfo {pages} {977} (\bibinfo {year} {2023})}\BibitemShut {NoStop}%
\bibitem [{\citenamefont {Zhao}\ \emph {et~al.}(2021)\citenamefont {Zhao}, \citenamefont {Pettes}, \citenamefont {Zheng},\ and\ \citenamefont {Htoon}}]{zhao2021site}%
  \BibitemOpen
  \bibfield  {author} {\bibinfo {author} {\bibfnamefont {H.}~\bibnamefont {Zhao}}, \bibinfo {author} {\bibfnamefont {M.~T.}\ \bibnamefont {Pettes}}, \bibinfo {author} {\bibfnamefont {Y.}~\bibnamefont {Zheng}},\ and\ \bibinfo {author} {\bibfnamefont {H.}~\bibnamefont {Htoon}},\ }\bibfield  {title} {\bibinfo {title} {Site-controlled telecom-wavelength single-photon emitters in atomically-thin mote2},\ }\href@noop {} {\bibfield  {journal} {\bibinfo  {journal} {Nat. Commun.}\ }\textbf {\bibinfo {volume} {12}},\ \bibinfo {pages} {6753} (\bibinfo {year} {2021})}\BibitemShut {NoStop}%
\bibitem [{\citenamefont {Borregaard}\ \emph {et~al.}(2019)\citenamefont {Borregaard}, \citenamefont {Sørensen},\ and\ \citenamefont {Lodahl}}]{borregaard2019quantum}%
  \BibitemOpen
  \bibfield  {author} {\bibinfo {author} {\bibfnamefont {J.}~\bibnamefont {Borregaard}}, \bibinfo {author} {\bibfnamefont {A.~S.}\ \bibnamefont {Sørensen}},\ and\ \bibinfo {author} {\bibfnamefont {P.}~\bibnamefont {Lodahl}},\ }\bibfield  {title} {\bibinfo {title} {Quantum networks with deterministic spin–photon interfaces},\ }\href@noop {} {\bibfield  {journal} {\bibinfo  {journal} {Adv. Quantum Technol.}\ }\textbf {\bibinfo {volume} {2}},\ \bibinfo {pages} {1800091} (\bibinfo {year} {2019})}\BibitemShut {NoStop}%
\bibitem [{\citenamefont {Kuhlmann}\ \emph {et~al.}(2015)\citenamefont {Kuhlmann}, \citenamefont {Prechtel}, \citenamefont {Houel}, \citenamefont {Ludwig}, \citenamefont {Reuter}, \citenamefont {Wieck},\ and\ \citenamefont {Warburton}}]{kuhlmann2015transform}%
  \BibitemOpen
  \bibfield  {author} {\bibinfo {author} {\bibfnamefont {A.~V.}\ \bibnamefont {Kuhlmann}}, \bibinfo {author} {\bibfnamefont {J.~H.}\ \bibnamefont {Prechtel}}, \bibinfo {author} {\bibfnamefont {J.}~\bibnamefont {Houel}}, \bibinfo {author} {\bibfnamefont {A.}~\bibnamefont {Ludwig}}, \bibinfo {author} {\bibfnamefont {D.}~\bibnamefont {Reuter}}, \bibinfo {author} {\bibfnamefont {A.~D.}\ \bibnamefont {Wieck}},\ and\ \bibinfo {author} {\bibfnamefont {R.~J.}\ \bibnamefont {Warburton}},\ }\bibfield  {title} {\bibinfo {title} {Transform-limited single photons from a single quantum dot},\ }\href@noop {} {\bibfield  {journal} {\bibinfo  {journal} {Nat. Commun.}\ }\textbf {\bibinfo {volume} {6}},\ \bibinfo {pages} {8204} (\bibinfo {year} {2015})}\BibitemShut {NoStop}%
\bibitem [{\citenamefont {Manga~Rao}\ and\ \citenamefont {Hughes}(2007)}]{manga2007single}%
  \BibitemOpen
  \bibfield  {author} {\bibinfo {author} {\bibfnamefont {V.}~\bibnamefont {Manga~Rao}}\ and\ \bibinfo {author} {\bibfnamefont {S.}~\bibnamefont {Hughes}},\ }\bibfield  {title} {\bibinfo {title} {Single quantum-dot purcell factor and $\beta$ factor in a photonic crystal waveguide},\ }\href@noop {} {\bibfield  {journal} {\bibinfo  {journal} {Phys. Rev. B}\ }\textbf {\bibinfo {volume} {75}},\ \bibinfo {pages} {205437} (\bibinfo {year} {2007})}\BibitemShut {NoStop}%
\bibitem [{\citenamefont {Warburton}(2013)}]{warburton2013single}%
  \BibitemOpen
  \bibfield  {author} {\bibinfo {author} {\bibfnamefont {R.~J.}\ \bibnamefont {Warburton}},\ }\bibfield  {title} {\bibinfo {title} {Single spins in self-assembled quantum dots},\ }\href@noop {} {\bibfield  {journal} {\bibinfo  {journal} {Nat. Mater.}\ }\textbf {\bibinfo {volume} {12}},\ \bibinfo {pages} {483} (\bibinfo {year} {2013})}\BibitemShut {NoStop}%
\bibitem [{\citenamefont {Kuhlmann}\ \emph {et~al.}(2013)\citenamefont {Kuhlmann}, \citenamefont {Houel}, \citenamefont {Ludwig}, \citenamefont {Greuter}, \citenamefont {Reuter}, \citenamefont {Wieck}, \citenamefont {Poggio},\ and\ \citenamefont {Warburton}}]{kuhlmann2013charge}%
  \BibitemOpen
  \bibfield  {author} {\bibinfo {author} {\bibfnamefont {A.~V.}\ \bibnamefont {Kuhlmann}}, \bibinfo {author} {\bibfnamefont {J.}~\bibnamefont {Houel}}, \bibinfo {author} {\bibfnamefont {A.}~\bibnamefont {Ludwig}}, \bibinfo {author} {\bibfnamefont {L.}~\bibnamefont {Greuter}}, \bibinfo {author} {\bibfnamefont {D.}~\bibnamefont {Reuter}}, \bibinfo {author} {\bibfnamefont {A.~D.}\ \bibnamefont {Wieck}}, \bibinfo {author} {\bibfnamefont {M.}~\bibnamefont {Poggio}},\ and\ \bibinfo {author} {\bibfnamefont {R.~J.}\ \bibnamefont {Warburton}},\ }\bibfield  {title} {\bibinfo {title} {Charge noise and spin noise in a semiconductor quantum device},\ }\href@noop {} {\bibfield  {journal} {\bibinfo  {journal} {Nat. Phys.}\ }\textbf {\bibinfo {volume} {9}},\ \bibinfo {pages} {570} (\bibinfo {year} {2013})}\BibitemShut {NoStop}%
\bibitem [{\citenamefont {Arcari}\ \emph {et~al.}(2014)\citenamefont {Arcari}, \citenamefont {S{\"o}llner}, \citenamefont {Javadi}, \citenamefont {Lindskov~Hansen}, \citenamefont {Mahmoodian}, \citenamefont {Liu}, \citenamefont {Thyrrestrup}, \citenamefont {Lee}, \citenamefont {Song}, \citenamefont {Stobbe} \emph {et~al.}}]{arcari2014near}%
  \BibitemOpen
  \bibfield  {author} {\bibinfo {author} {\bibfnamefont {M.}~\bibnamefont {Arcari}}, \bibinfo {author} {\bibfnamefont {I.}~\bibnamefont {S{\"o}llner}}, \bibinfo {author} {\bibfnamefont {A.}~\bibnamefont {Javadi}}, \bibinfo {author} {\bibfnamefont {S.}~\bibnamefont {Lindskov~Hansen}}, \bibinfo {author} {\bibfnamefont {S.}~\bibnamefont {Mahmoodian}}, \bibinfo {author} {\bibfnamefont {J.}~\bibnamefont {Liu}}, \bibinfo {author} {\bibfnamefont {H.}~\bibnamefont {Thyrrestrup}}, \bibinfo {author} {\bibfnamefont {E.~H.}\ \bibnamefont {Lee}}, \bibinfo {author} {\bibfnamefont {J.~D.}\ \bibnamefont {Song}}, \bibinfo {author} {\bibfnamefont {S.}~\bibnamefont {Stobbe}}, \emph {et~al.},\ }\bibfield  {title} {\bibinfo {title} {Near-unity coupling efficiency of a quantum emitter to a photonic crystal waveguide},\ }\href@noop {} {\bibfield  {journal} {\bibinfo  {journal} {Phys. Rev. Lett.}\ }\textbf {\bibinfo {volume} {113}},\ \bibinfo {pages} {093603} (\bibinfo {year} {2014})}\BibitemShut {NoStop}%
\bibitem [{\citenamefont {Zhou}\ \emph {et~al.}(2018)\citenamefont {Zhou}, \citenamefont {Kulkova}, \citenamefont {Lund-Hansen}, \citenamefont {Lindskov~Hansen}, \citenamefont {Lodahl},\ and\ \citenamefont {Midolo}}]{zhou2018high}%
  \BibitemOpen
  \bibfield  {author} {\bibinfo {author} {\bibfnamefont {X.}~\bibnamefont {Zhou}}, \bibinfo {author} {\bibfnamefont {I.}~\bibnamefont {Kulkova}}, \bibinfo {author} {\bibfnamefont {T.}~\bibnamefont {Lund-Hansen}}, \bibinfo {author} {\bibfnamefont {S.}~\bibnamefont {Lindskov~Hansen}}, \bibinfo {author} {\bibfnamefont {P.}~\bibnamefont {Lodahl}},\ and\ \bibinfo {author} {\bibfnamefont {L.}~\bibnamefont {Midolo}},\ }\bibfield  {title} {\bibinfo {title} {High-efficiency shallow-etched grating on gaas membranes for quantum photonic applications},\ }\href@noop {} {\bibfield  {journal} {\bibinfo  {journal} {Appl. Phys. Lett.}\ }\textbf {\bibinfo {volume} {113}},\ \bibinfo {pages} {251103} (\bibinfo {year} {2018})}\BibitemShut {NoStop}%
\bibitem [{\citenamefont {Wang}\ \emph {et~al.}(2021)\citenamefont {Wang}, \citenamefont {Uppu}, \citenamefont {Zhou}, \citenamefont {Papon}, \citenamefont {Scholz}, \citenamefont {Wieck}, \citenamefont {Ludwig}, \citenamefont {Lodahl},\ and\ \citenamefont {Midolo}}]{wang2021electroabsorption}%
  \BibitemOpen
  \bibfield  {author} {\bibinfo {author} {\bibfnamefont {Y.}~\bibnamefont {Wang}}, \bibinfo {author} {\bibfnamefont {R.}~\bibnamefont {Uppu}}, \bibinfo {author} {\bibfnamefont {X.}~\bibnamefont {Zhou}}, \bibinfo {author} {\bibfnamefont {C.}~\bibnamefont {Papon}}, \bibinfo {author} {\bibfnamefont {S.}~\bibnamefont {Scholz}}, \bibinfo {author} {\bibfnamefont {A.~D.}\ \bibnamefont {Wieck}}, \bibinfo {author} {\bibfnamefont {A.}~\bibnamefont {Ludwig}}, \bibinfo {author} {\bibfnamefont {P.}~\bibnamefont {Lodahl}},\ and\ \bibinfo {author} {\bibfnamefont {L.}~\bibnamefont {Midolo}},\ }\bibfield  {title} {\bibinfo {title} {Electroabsorption in gated gaas nanophotonic waveguides},\ }\href@noop {} {\bibfield  {journal} {\bibinfo  {journal} {Appl. Phys. Lett.}\ }\textbf {\bibinfo {volume} {118}},\ \bibinfo {pages} {131106} (\bibinfo {year} {2021})}\BibitemShut {NoStop}%
\bibitem [{\citenamefont {Papon}\ \emph {et~al.}(2023)\citenamefont {Papon}, \citenamefont {Wang}, \citenamefont {Uppu}, \citenamefont {Scholz}, \citenamefont {Wieck}, \citenamefont {Ludwig}, \citenamefont {Lodahl},\ and\ \citenamefont {Midolo}}]{papon2023independent}%
  \BibitemOpen
  \bibfield  {author} {\bibinfo {author} {\bibfnamefont {C.}~\bibnamefont {Papon}}, \bibinfo {author} {\bibfnamefont {Y.}~\bibnamefont {Wang}}, \bibinfo {author} {\bibfnamefont {R.}~\bibnamefont {Uppu}}, \bibinfo {author} {\bibfnamefont {S.}~\bibnamefont {Scholz}}, \bibinfo {author} {\bibfnamefont {A.~D.}\ \bibnamefont {Wieck}}, \bibinfo {author} {\bibfnamefont {A.}~\bibnamefont {Ludwig}}, \bibinfo {author} {\bibfnamefont {P.}~\bibnamefont {Lodahl}},\ and\ \bibinfo {author} {\bibfnamefont {L.}~\bibnamefont {Midolo}},\ }\bibfield  {title} {\bibinfo {title} {Independent operation of two waveguide-integrated quantum emitters},\ }\href@noop {} {\bibfield  {journal} {\bibinfo  {journal} {Phys. Rev. Appl.}\ }\textbf {\bibinfo {volume} {19}},\ \bibinfo {pages} {L061003} (\bibinfo {year} {2023})}\BibitemShut {NoStop}%
\bibitem [{\citenamefont {Tiranov}\ \emph {et~al.}(2023)\citenamefont {Tiranov}, \citenamefont {Lodahl} \emph {et~al.}}]{tiranov2023collective}%
  \BibitemOpen
  \bibfield  {author} {\bibinfo {author} {\bibfnamefont {A.}~\bibnamefont {Tiranov}}, \bibinfo {author} {\bibfnamefont {P.}~\bibnamefont {Lodahl}}, \emph {et~al.},\ }\bibfield  {title} {\bibinfo {title} {Collective super- and subradiant dynamics between distant quantum emitters},\ }\href@noop {} {\bibfield  {journal} {\bibinfo  {journal} {Science}\ }\textbf {\bibinfo {volume} {379}},\ \bibinfo {pages} {389} (\bibinfo {year} {2023})}\BibitemShut {NoStop}%
\bibitem [{\citenamefont {Zahidy}\ \emph {et~al.}(2024)\citenamefont {Zahidy}, \citenamefont {Mikkelsen}, \citenamefont {M{\"u}ller}, \citenamefont {Da~Lio}, \citenamefont {Krehbiel}, \citenamefont {Wang}, \citenamefont {Bart}, \citenamefont {Wieck}, \citenamefont {Ludwig}, \citenamefont {Galili} \emph {et~al.}}]{zahidy2024quantum}%
  \BibitemOpen
  \bibfield  {author} {\bibinfo {author} {\bibfnamefont {M.}~\bibnamefont {Zahidy}}, \bibinfo {author} {\bibfnamefont {M.~T.}\ \bibnamefont {Mikkelsen}}, \bibinfo {author} {\bibfnamefont {R.}~\bibnamefont {M{\"u}ller}}, \bibinfo {author} {\bibfnamefont {B.}~\bibnamefont {Da~Lio}}, \bibinfo {author} {\bibfnamefont {M.}~\bibnamefont {Krehbiel}}, \bibinfo {author} {\bibfnamefont {Y.}~\bibnamefont {Wang}}, \bibinfo {author} {\bibfnamefont {N.}~\bibnamefont {Bart}}, \bibinfo {author} {\bibfnamefont {A.~D.}\ \bibnamefont {Wieck}}, \bibinfo {author} {\bibfnamefont {A.}~\bibnamefont {Ludwig}}, \bibinfo {author} {\bibfnamefont {M.}~\bibnamefont {Galili}}, \emph {et~al.},\ }\bibfield  {title} {\bibinfo {title} {Quantum key distribution using deterministic single-photon sources over a field-installed fibre link},\ }\href@noop {} {\bibfield  {journal} {\bibinfo  {journal} {npj Quantum Inf.}\ }\textbf {\bibinfo {volume} {10}},\ \bibinfo {pages} {2} (\bibinfo {year} {2024})}\BibitemShut {NoStop}%
\bibitem [{\citenamefont {Nawrath}\ \emph {et~al.}(2019)\citenamefont {Nawrath}, \citenamefont {Olbrich}, \citenamefont {Paul}, \citenamefont {Portalupi}, \citenamefont {Jetter},\ and\ \citenamefont {Michler}}]{nawrath2019coherence}%
  \BibitemOpen
  \bibfield  {author} {\bibinfo {author} {\bibfnamefont {C.}~\bibnamefont {Nawrath}}, \bibinfo {author} {\bibfnamefont {F.}~\bibnamefont {Olbrich}}, \bibinfo {author} {\bibfnamefont {M.}~\bibnamefont {Paul}}, \bibinfo {author} {\bibfnamefont {S.}~\bibnamefont {Portalupi}}, \bibinfo {author} {\bibfnamefont {M.}~\bibnamefont {Jetter}},\ and\ \bibinfo {author} {\bibfnamefont {P.}~\bibnamefont {Michler}},\ }\bibfield  {title} {\bibinfo {title} {Coherence and indistinguishability of highly pure single photons from non-resonantly and resonantly excited telecom c-band quantum dots},\ }\href@noop {} {\bibfield  {journal} {\bibinfo  {journal} {Appl. Phys. Lett.}\ }\textbf {\bibinfo {volume} {115}} (\bibinfo {year} {2019})}\BibitemShut {NoStop}%
\bibitem [{\citenamefont {Hauser}\ \emph {et~al.}(2025)\citenamefont {Hauser}, \citenamefont {Bayerbach}, \citenamefont {Kaupp}, \citenamefont {Reum}, \citenamefont {Peniakov}, \citenamefont {Michl}, \citenamefont {Kamp}, \citenamefont {Huber-Loyola}, \citenamefont {Pfenning}, \citenamefont {H{\"o}fling} \emph {et~al.}}]{hauser2025deterministic}%
  \BibitemOpen
  \bibfield  {author} {\bibinfo {author} {\bibfnamefont {N.}~\bibnamefont {Hauser}}, \bibinfo {author} {\bibfnamefont {M.}~\bibnamefont {Bayerbach}}, \bibinfo {author} {\bibfnamefont {J.}~\bibnamefont {Kaupp}}, \bibinfo {author} {\bibfnamefont {Y.}~\bibnamefont {Reum}}, \bibinfo {author} {\bibfnamefont {G.}~\bibnamefont {Peniakov}}, \bibinfo {author} {\bibfnamefont {J.}~\bibnamefont {Michl}}, \bibinfo {author} {\bibfnamefont {M.}~\bibnamefont {Kamp}}, \bibinfo {author} {\bibfnamefont {T.}~\bibnamefont {Huber-Loyola}}, \bibinfo {author} {\bibfnamefont {A.~T.}\ \bibnamefont {Pfenning}}, \bibinfo {author} {\bibfnamefont {S.}~\bibnamefont {H{\"o}fling}}, \emph {et~al.},\ }\bibfield  {title} {\bibinfo {title} {Deterministic and highly indistinguishable single photons in the telecom c-band},\ }\href@noop {} {\bibfield  {journal} {\bibinfo  {journal} {Preprint at arXiv:2505.09695}\ } (\bibinfo {year} {2025})}\BibitemShut {NoStop}%
\bibitem [{\citenamefont {Holewa}\ \emph {et~al.}(2024)\citenamefont {Holewa}, \citenamefont {Vajner}, \citenamefont {Z.}, \citenamefont {Wasiluk}, \citenamefont {Ga{\'a}l}, \citenamefont {Sakanas}, \citenamefont {Burakowski}, \citenamefont {Mrowi{\'n}ski}, \citenamefont {Krajnik}, \citenamefont {Xiong} \emph {et~al.}}]{holewa2024high}%
  \BibitemOpen
  \bibfield  {author} {\bibinfo {author} {\bibfnamefont {P.}~\bibnamefont {Holewa}}, \bibinfo {author} {\bibfnamefont {D.~A.}\ \bibnamefont {Vajner}}, \bibinfo {author} {\bibfnamefont {E.}~\bibnamefont {Z.}}, \bibinfo {author} {\bibfnamefont {M.}~\bibnamefont {Wasiluk}}, \bibinfo {author} {\bibfnamefont {B.}~\bibnamefont {Ga{\'a}l}}, \bibinfo {author} {\bibfnamefont {A.}~\bibnamefont {Sakanas}}, \bibinfo {author} {\bibfnamefont {M.}~\bibnamefont {Burakowski}}, \bibinfo {author} {\bibfnamefont {P.}~\bibnamefont {Mrowi{\'n}ski}}, \bibinfo {author} {\bibfnamefont {B.}~\bibnamefont {Krajnik}}, \bibinfo {author} {\bibfnamefont {M.}~\bibnamefont {Xiong}}, \emph {et~al.},\ }\bibfield  {title} {\bibinfo {title} {High-throughput quantum photonic devices emitting indistinguishable photons in the telecom c-band},\ }\href@noop {} {\bibfield  {journal} {\bibinfo  {journal} {Nat. Commun.}\ }\textbf {\bibinfo {volume} {15}},\ \bibinfo {pages} {3358} (\bibinfo {year} {2024})}\BibitemShut {NoStop}%
\bibitem [{\citenamefont {Aghaee~Rad}\ \emph {et~al.}(2025)\citenamefont {Aghaee~Rad}, \citenamefont {Ainsworth}, \citenamefont {Alexander}, \citenamefont {Altieri}, \citenamefont {Askarani}, \citenamefont {Baby}, \citenamefont {Banchi}, \citenamefont {Baragiola}, \citenamefont {Bourassa}, \citenamefont {Chadwick} \emph {et~al.}}]{aghaee2025scaling}%
  \BibitemOpen
  \bibfield  {author} {\bibinfo {author} {\bibfnamefont {H.}~\bibnamefont {Aghaee~Rad}}, \bibinfo {author} {\bibfnamefont {T.}~\bibnamefont {Ainsworth}}, \bibinfo {author} {\bibfnamefont {R.}~\bibnamefont {Alexander}}, \bibinfo {author} {\bibfnamefont {B.}~\bibnamefont {Altieri}}, \bibinfo {author} {\bibfnamefont {M.}~\bibnamefont {Askarani}}, \bibinfo {author} {\bibfnamefont {R.}~\bibnamefont {Baby}}, \bibinfo {author} {\bibfnamefont {L.}~\bibnamefont {Banchi}}, \bibinfo {author} {\bibfnamefont {B.}~\bibnamefont {Baragiola}}, \bibinfo {author} {\bibfnamefont {J.}~\bibnamefont {Bourassa}}, \bibinfo {author} {\bibfnamefont {R.}~\bibnamefont {Chadwick}}, \emph {et~al.},\ }\bibfield  {title} {\bibinfo {title} {Scaling and networking a modular photonic quantum computer},\ }\href@noop {} {\bibfield  {journal} {\bibinfo  {journal} {Nature}\ }\textbf {\bibinfo {volume} {638}},\ \bibinfo {pages} {912} (\bibinfo {year} {2025})}\BibitemShut {NoStop}%
\bibitem [{\citenamefont {Da~Lio}\ \emph {et~al.}(2022)\citenamefont {Da~Lio}, \citenamefont {Faurby}, \citenamefont {Zhou}, \citenamefont {Chan}, \citenamefont {Uppu}, \citenamefont {Thyrrestrup}, \citenamefont {Scholz}, \citenamefont {Wieck}, \citenamefont {Ludwig}, \citenamefont {Lodahl},\ and\ \citenamefont {Midolo}}]{da_lio2022pure}%
  \BibitemOpen
  \bibfield  {author} {\bibinfo {author} {\bibfnamefont {B.}~\bibnamefont {Da~Lio}}, \bibinfo {author} {\bibfnamefont {C.}~\bibnamefont {Faurby}}, \bibinfo {author} {\bibfnamefont {X.}~\bibnamefont {Zhou}}, \bibinfo {author} {\bibfnamefont {M.~L.}\ \bibnamefont {Chan}}, \bibinfo {author} {\bibfnamefont {R.}~\bibnamefont {Uppu}}, \bibinfo {author} {\bibfnamefont {H.}~\bibnamefont {Thyrrestrup}}, \bibinfo {author} {\bibfnamefont {S.}~\bibnamefont {Scholz}}, \bibinfo {author} {\bibfnamefont {A.~D.}\ \bibnamefont {Wieck}}, \bibinfo {author} {\bibfnamefont {A.}~\bibnamefont {Ludwig}}, \bibinfo {author} {\bibfnamefont {P.}~\bibnamefont {Lodahl}},\ and\ \bibinfo {author} {\bibfnamefont {L.}~\bibnamefont {Midolo}},\ }\bibfield  {title} {\bibinfo {title} {A pure and indistinguishable single-photon source at telecommunication wavelength},\ }\href@noop {} {\bibfield  {journal} {\bibinfo  {journal} {Adv. Quantum Technol.}\ }\textbf {\bibinfo {volume} {5}},\ \bibinfo {pages} {2200006} (\bibinfo {year} {2022})}\BibitemShut
  {NoStop}%
\bibitem [{\citenamefont {Kurzmann}\ \emph {et~al.}(2016)\citenamefont {Kurzmann}, \citenamefont {Ludwig}, \citenamefont {Wieck}, \citenamefont {Lorke},\ and\ \citenamefont {Geller}}]{kurzmann2016auger}%
  \BibitemOpen
  \bibfield  {author} {\bibinfo {author} {\bibfnamefont {A.}~\bibnamefont {Kurzmann}}, \bibinfo {author} {\bibfnamefont {A.}~\bibnamefont {Ludwig}}, \bibinfo {author} {\bibfnamefont {A.~D.}\ \bibnamefont {Wieck}}, \bibinfo {author} {\bibfnamefont {A.}~\bibnamefont {Lorke}},\ and\ \bibinfo {author} {\bibfnamefont {M.}~\bibnamefont {Geller}},\ }\bibfield  {title} {\bibinfo {title} {Auger recombination in self-assembled quantum dots: quenching and broadening of the charged exciton transition},\ }\href@noop {} {\bibfield  {journal} {\bibinfo  {journal} {Nano Lett.}\ }\textbf {\bibinfo {volume} {16}},\ \bibinfo {pages} {3367} (\bibinfo {year} {2016})}\BibitemShut {NoStop}%
\bibitem [{\citenamefont {Sund}\ \emph {et~al.}(2023)\citenamefont {Sund}, \citenamefont {Lenzini}, \citenamefont {Paesani} \emph {et~al.}}]{sund2023high}%
  \BibitemOpen
  \bibfield  {author} {\bibinfo {author} {\bibfnamefont {P.~I.}\ \bibnamefont {Sund}}, \bibinfo {author} {\bibfnamefont {F.}~\bibnamefont {Lenzini}}, \bibinfo {author} {\bibfnamefont {S.}~\bibnamefont {Paesani}}, \emph {et~al.},\ }\bibfield  {title} {\bibinfo {title} {High-speed thin-film lithium niobate quantum processor driven by a solid-state quantum emitter},\ }\href@noop {} {\bibfield  {journal} {\bibinfo  {journal} {Sci. Adv.}\ }\textbf {\bibinfo {volume} {9}},\ \bibinfo {pages} {eadg7268} (\bibinfo {year} {2023})}\BibitemShut {NoStop}%
\bibitem [{\citenamefont {Gonz{\'a}lez-Ruiz}\ \emph {et~al.}(2025)\citenamefont {Gonz{\'a}lez-Ruiz}, \citenamefont {Bjerlin}, \citenamefont {Sandberg},\ and\ \citenamefont {S{\o}rensen}}]{gonzalez2025two}%
  \BibitemOpen
  \bibfield  {author} {\bibinfo {author} {\bibfnamefont {E.~M.}\ \bibnamefont {Gonz{\'a}lez-Ruiz}}, \bibinfo {author} {\bibfnamefont {J.}~\bibnamefont {Bjerlin}}, \bibinfo {author} {\bibfnamefont {O.~A.~D.}\ \bibnamefont {Sandberg}},\ and\ \bibinfo {author} {\bibfnamefont {A.~S.}\ \bibnamefont {S{\o}rensen}},\ }\bibfield  {title} {\bibinfo {title} {Two-photon correlations and hong-ou-mandel visibility from an imperfect single-photon source},\ }\href@noop {} {\bibfield  {journal} {\bibinfo  {journal} {Phys. Rev. Appl.}\ }\textbf {\bibinfo {volume} {23}},\ \bibinfo {pages} {054063} (\bibinfo {year} {2025})}\BibitemShut {NoStop}%
\bibitem [{\citenamefont {Zhang}\ \emph {et~al.}(2019)\citenamefont {Zhang}, \citenamefont {Muliuk}, \citenamefont {Juvert}, \citenamefont {Kumari}, \citenamefont {Goyvaerts}, \citenamefont {Haq}, \citenamefont {Op~de Beeck}, \citenamefont {Kuyken}, \citenamefont {Morthier}, \citenamefont {Van~Thourhout} \emph {et~al.}}]{zhang2019iii}%
  \BibitemOpen
  \bibfield  {author} {\bibinfo {author} {\bibfnamefont {J.}~\bibnamefont {Zhang}}, \bibinfo {author} {\bibfnamefont {G.}~\bibnamefont {Muliuk}}, \bibinfo {author} {\bibfnamefont {J.}~\bibnamefont {Juvert}}, \bibinfo {author} {\bibfnamefont {S.}~\bibnamefont {Kumari}}, \bibinfo {author} {\bibfnamefont {J.}~\bibnamefont {Goyvaerts}}, \bibinfo {author} {\bibfnamefont {B.}~\bibnamefont {Haq}}, \bibinfo {author} {\bibfnamefont {C.}~\bibnamefont {Op~de Beeck}}, \bibinfo {author} {\bibfnamefont {B.}~\bibnamefont {Kuyken}}, \bibinfo {author} {\bibfnamefont {G.}~\bibnamefont {Morthier}}, \bibinfo {author} {\bibfnamefont {D.}~\bibnamefont {Van~Thourhout}}, \emph {et~al.},\ }\bibfield  {title} {\bibinfo {title} {Iii-v-on-si photonic integrated circuits realized using micro-transfer-printing},\ }\href@noop {} {\bibfield  {journal} {\bibinfo  {journal} {APL Photonics}\ }\textbf {\bibinfo {volume} {4}} (\bibinfo {year} {2019})}\BibitemShut {NoStop}%
\bibitem [{\citenamefont {Davanco}\ \emph {et~al.}(2017)\citenamefont {Davanco}, \citenamefont {Liu}, \citenamefont {Sapienza}, \citenamefont {Zhang}, \citenamefont {De~Miranda~Cardoso}, \citenamefont {Verma}, \citenamefont {Mirin}, \citenamefont {Nam}, \citenamefont {Liu},\ and\ \citenamefont {Srinivasan}}]{davanco2017heterogeneous}%
  \BibitemOpen
  \bibfield  {author} {\bibinfo {author} {\bibfnamefont {M.}~\bibnamefont {Davanco}}, \bibinfo {author} {\bibfnamefont {J.}~\bibnamefont {Liu}}, \bibinfo {author} {\bibfnamefont {L.}~\bibnamefont {Sapienza}}, \bibinfo {author} {\bibfnamefont {C.-Z.}\ \bibnamefont {Zhang}}, \bibinfo {author} {\bibfnamefont {J.~V.}\ \bibnamefont {De~Miranda~Cardoso}}, \bibinfo {author} {\bibfnamefont {V.}~\bibnamefont {Verma}}, \bibinfo {author} {\bibfnamefont {R.}~\bibnamefont {Mirin}}, \bibinfo {author} {\bibfnamefont {S.~W.}\ \bibnamefont {Nam}}, \bibinfo {author} {\bibfnamefont {L.}~\bibnamefont {Liu}},\ and\ \bibinfo {author} {\bibfnamefont {K.}~\bibnamefont {Srinivasan}},\ }\bibfield  {title} {\bibinfo {title} {Heterogeneous integration for on-chip quantum photonic circuits with single quantum dot devices},\ }\href@noop {} {\bibfield  {journal} {\bibinfo  {journal} {Nat. Commun.}\ }\textbf {\bibinfo {volume} {8}},\ \bibinfo {pages} {889} (\bibinfo {year} {2017})}\BibitemShut {NoStop}%
\bibitem [{\citenamefont {Salamon}\ \emph {et~al.}(2025)\citenamefont {Salamon}, \citenamefont {Wang}, \citenamefont {Snedker-Nielsen}, \citenamefont {Shadmani}, \citenamefont {Schott}, \citenamefont {Balauroiu}, \citenamefont {Volet}, \citenamefont {Ludwig},\ and\ \citenamefont {Midolo}}]{salamon2025electrical}%
  \BibitemOpen
  \bibfield  {author} {\bibinfo {author} {\bibfnamefont {H.}~\bibnamefont {Salamon}}, \bibinfo {author} {\bibfnamefont {Y.}~\bibnamefont {Wang}}, \bibinfo {author} {\bibfnamefont {A.}~\bibnamefont {Snedker-Nielsen}}, \bibinfo {author} {\bibfnamefont {A.}~\bibnamefont {Shadmani}}, \bibinfo {author} {\bibfnamefont {R.}~\bibnamefont {Schott}}, \bibinfo {author} {\bibfnamefont {M.}~\bibnamefont {Balauroiu}}, \bibinfo {author} {\bibfnamefont {N.}~\bibnamefont {Volet}}, \bibinfo {author} {\bibfnamefont {A.}~\bibnamefont {Ludwig}},\ and\ \bibinfo {author} {\bibfnamefont {L.}~\bibnamefont {Midolo}},\ }\bibfield  {title} {\bibinfo {title} {Electrical control of quantum dots in gaas-on-insulator waveguides for coherent single-photon generation},\ }\href@noop {} {\bibfield  {journal} {\bibinfo  {journal} {Preprint at arXiv:2508.04584}\ } (\bibinfo {year} {2025})}\BibitemShut {NoStop}%
\bibitem [{\citenamefont {Bernal}\ \emph {et~al.}(2024)\citenamefont {Bernal}, \citenamefont {Dumont}, \citenamefont {Berikaa}, \citenamefont {St-Arnault}, \citenamefont {Hu}, \citenamefont {Castrejon}, \citenamefont {Li}, \citenamefont {Wei}, \citenamefont {Krueger}, \citenamefont {Pittal{\`a}} \emph {et~al.}}]{bernal202412}%
  \BibitemOpen
  \bibfield  {author} {\bibinfo {author} {\bibfnamefont {S.}~\bibnamefont {Bernal}}, \bibinfo {author} {\bibfnamefont {M.}~\bibnamefont {Dumont}}, \bibinfo {author} {\bibfnamefont {E.}~\bibnamefont {Berikaa}}, \bibinfo {author} {\bibfnamefont {C.}~\bibnamefont {St-Arnault}}, \bibinfo {author} {\bibfnamefont {Y.}~\bibnamefont {Hu}}, \bibinfo {author} {\bibfnamefont {R.~G.}\ \bibnamefont {Castrejon}}, \bibinfo {author} {\bibfnamefont {W.}~\bibnamefont {Li}}, \bibinfo {author} {\bibfnamefont {Z.}~\bibnamefont {Wei}}, \bibinfo {author} {\bibfnamefont {B.}~\bibnamefont {Krueger}}, \bibinfo {author} {\bibfnamefont {F.}~\bibnamefont {Pittal{\`a}}}, \emph {et~al.},\ }\bibfield  {title} {\bibinfo {title} {12.1 terabit/second data center interconnects using o-band coherent transmission with qd-mll frequency combs},\ }\href@noop {} {\bibfield  {journal} {\bibinfo  {journal} {Nat. Commun.}\ }\textbf {\bibinfo {volume} {15}},\ \bibinfo {pages} {7741} (\bibinfo {year} {2024})}\BibitemShut {NoStop}%
\bibitem [{\citenamefont {Ottaviano}\ \emph {et~al.}(2016)\citenamefont {Ottaviano}, \citenamefont {Pu}, \citenamefont {Semenova},\ and\ \citenamefont {Yvind}}]{ottaviano2016}%
  \BibitemOpen
  \bibfield  {author} {\bibinfo {author} {\bibfnamefont {L.}~\bibnamefont {Ottaviano}}, \bibinfo {author} {\bibfnamefont {M.}~\bibnamefont {Pu}}, \bibinfo {author} {\bibfnamefont {E.}~\bibnamefont {Semenova}},\ and\ \bibinfo {author} {\bibfnamefont {K.}~\bibnamefont {Yvind}},\ }\bibfield  {title} {\bibinfo {title} {Low‐loss high‐confinement waveguides and microring resonators in algaas‐on‐insulator},\ }\href@noop {} {\bibfield  {journal} {\bibinfo  {journal} {Opt. Lett.}\ }\textbf {\bibinfo {volume} {41}},\ \bibinfo {pages} {3996} (\bibinfo {year} {2016})}\BibitemShut {NoStop}%
\bibitem [{\citenamefont {Chang}\ \emph {et~al.}(2020)\citenamefont {Chang}, \citenamefont {Xie}, \citenamefont {Shu}, \citenamefont {Yang}, \citenamefont {Shen}, \citenamefont {Boes}, \citenamefont {Peters}, \citenamefont {Jin}, \citenamefont {Xiang}, \citenamefont {Liu}, \citenamefont {Moille}, \citenamefont {Yu}, \citenamefont {Wang}, \citenamefont {Srinivasan}, \citenamefont {Papp}, \citenamefont {Vahala},\ and\ \citenamefont {Bowers}}]{chang2020}%
  \BibitemOpen
  \bibfield  {author} {\bibinfo {author} {\bibfnamefont {L.}~\bibnamefont {Chang}}, \bibinfo {author} {\bibfnamefont {W.}~\bibnamefont {Xie}}, \bibinfo {author} {\bibfnamefont {H.}~\bibnamefont {Shu}}, \bibinfo {author} {\bibfnamefont {Q.-F.}\ \bibnamefont {Yang}}, \bibinfo {author} {\bibfnamefont {B.}~\bibnamefont {Shen}}, \bibinfo {author} {\bibfnamefont {A.}~\bibnamefont {Boes}}, \bibinfo {author} {\bibfnamefont {J.~D.}\ \bibnamefont {Peters}}, \bibinfo {author} {\bibfnamefont {W.}~\bibnamefont {Jin}}, \bibinfo {author} {\bibfnamefont {C.}~\bibnamefont {Xiang}}, \bibinfo {author} {\bibfnamefont {S.}~\bibnamefont {Liu}}, \bibinfo {author} {\bibfnamefont {G.}~\bibnamefont {Moille}}, \bibinfo {author} {\bibfnamefont {S.-P.}\ \bibnamefont {Yu}}, \bibinfo {author} {\bibfnamefont {X.}~\bibnamefont {Wang}}, \bibinfo {author} {\bibfnamefont {K.}~\bibnamefont {Srinivasan}}, \bibinfo {author} {\bibfnamefont {S.~B.}\ \bibnamefont {Papp}}, \bibinfo {author} {\bibfnamefont {K.}~\bibnamefont {Vahala}},\ and\ \bibinfo
  {author} {\bibfnamefont {J.~E.}\ \bibnamefont {Bowers}},\ }\bibfield  {title} {\bibinfo {title} {Ultra-efficient frequency comb generation in algaas-on-insulator microresonators},\ }\href@noop {} {\bibfield  {journal} {\bibinfo  {journal} {Nat. Commun.}\ }\textbf {\bibinfo {volume} {11}},\ \bibinfo {pages} {1331} (\bibinfo {year} {2020})}\BibitemShut {NoStop}%
\bibitem [{\citenamefont {Sprengers}\ \emph {et~al.}(2011)\citenamefont {Sprengers}, \citenamefont {Gaggero}, \citenamefont {Sahin}, \citenamefont {Jahanmirinejad}, \citenamefont {Frucci}, \citenamefont {Mattioli}, \citenamefont {Leoni}, \citenamefont {Beetz}, \citenamefont {Lermer}, \citenamefont {Kamp} \emph {et~al.}}]{sprengers2011waveguide}%
  \BibitemOpen
  \bibfield  {author} {\bibinfo {author} {\bibfnamefont {J.}~\bibnamefont {Sprengers}}, \bibinfo {author} {\bibfnamefont {A.}~\bibnamefont {Gaggero}}, \bibinfo {author} {\bibfnamefont {D.}~\bibnamefont {Sahin}}, \bibinfo {author} {\bibfnamefont {S.}~\bibnamefont {Jahanmirinejad}}, \bibinfo {author} {\bibfnamefont {G.}~\bibnamefont {Frucci}}, \bibinfo {author} {\bibfnamefont {F.}~\bibnamefont {Mattioli}}, \bibinfo {author} {\bibfnamefont {R.}~\bibnamefont {Leoni}}, \bibinfo {author} {\bibfnamefont {J.}~\bibnamefont {Beetz}}, \bibinfo {author} {\bibfnamefont {M.}~\bibnamefont {Lermer}}, \bibinfo {author} {\bibfnamefont {M.}~\bibnamefont {Kamp}}, \emph {et~al.},\ }\bibfield  {title} {\bibinfo {title} {Waveguide superconducting single-photon detectors for integrated quantum photonic circuits},\ }\href@noop {} {\bibfield  {journal} {\bibinfo  {journal} {Appl. Phys. Lett.}\ }\textbf {\bibinfo {volume} {99}} (\bibinfo {year} {2011})}\BibitemShut {NoStop}%
\bibitem [{\citenamefont {Nishi}\ \emph {et~al.}(1999)\citenamefont {Nishi}, \citenamefont {Saito}, \citenamefont {Sugou},\ and\ \citenamefont {Lee}}]{NISHI1999}%
  \BibitemOpen
  \bibfield  {author} {\bibinfo {author} {\bibfnamefont {K.}~\bibnamefont {Nishi}}, \bibinfo {author} {\bibfnamefont {H.}~\bibnamefont {Saito}}, \bibinfo {author} {\bibfnamefont {S.}~\bibnamefont {Sugou}},\ and\ \bibinfo {author} {\bibfnamefont {J.-S.}\ \bibnamefont {Lee}},\ }\bibfield  {title} {\bibinfo {title} {A narrow photoluminescence linewidth of 21 mev at {1.35 \textmu m} from strain-reduced inas quantum dots covered by in0.2ga0.8as grown on gaas substrates},\ }\href@noop {} {\bibfield  {journal} {\bibinfo  {journal} {Appl. Phys. Lett.}\ }\textbf {\bibinfo {volume} {74}},\ \bibinfo {pages} {1111} (\bibinfo {year} {1999})}\BibitemShut {NoStop}%
\bibitem [{\citenamefont {Ludwig}\ \emph {et~al.}(2017)\citenamefont {Ludwig}, \citenamefont {Prechtel}, \citenamefont {Kuhlmann}, \citenamefont {Houel}, \citenamefont {Valentin}, \citenamefont {Warburton},\ and\ \citenamefont {Wieck}}]{LUDWIG2017}%
  \BibitemOpen
  \bibfield  {author} {\bibinfo {author} {\bibfnamefont {A.}~\bibnamefont {Ludwig}}, \bibinfo {author} {\bibfnamefont {J.~H.}\ \bibnamefont {Prechtel}}, \bibinfo {author} {\bibfnamefont {A.~V.}\ \bibnamefont {Kuhlmann}}, \bibinfo {author} {\bibfnamefont {J.}~\bibnamefont {Houel}}, \bibinfo {author} {\bibfnamefont {S.~R.}\ \bibnamefont {Valentin}}, \bibinfo {author} {\bibfnamefont {R.~J.}\ \bibnamefont {Warburton}},\ and\ \bibinfo {author} {\bibfnamefont {A.~D.}\ \bibnamefont {Wieck}},\ }\bibfield  {title} {\bibinfo {title} {Ultra-low charge and spin noise in self-assembled quantum dots},\ }\href@noop {} {\bibfield  {journal} {\bibinfo  {journal} {J. Cryst. Growth}\ }\textbf {\bibinfo {volume} {477}},\ \bibinfo {pages} {193} (\bibinfo {year} {2017})}\BibitemShut {NoStop}%
\bibitem [{\citenamefont {Nguyen}\ \emph {et~al.}(2020)\citenamefont {Nguyen}, \citenamefont {Korsch}, \citenamefont {Schmidt}, \citenamefont {Ebler}, \citenamefont {Labud}, \citenamefont {Schott}, \citenamefont {Lochner}, \citenamefont {Brinks}, \citenamefont {Wieck},\ and\ \citenamefont {Ludwig}}]{NGUYEN2020}%
  \BibitemOpen
  \bibfield  {author} {\bibinfo {author} {\bibfnamefont {G.}~\bibnamefont {Nguyen}}, \bibinfo {author} {\bibfnamefont {A.}~\bibnamefont {Korsch}}, \bibinfo {author} {\bibfnamefont {M.}~\bibnamefont {Schmidt}}, \bibinfo {author} {\bibfnamefont {C.}~\bibnamefont {Ebler}}, \bibinfo {author} {\bibfnamefont {P.}~\bibnamefont {Labud}}, \bibinfo {author} {\bibfnamefont {R.}~\bibnamefont {Schott}}, \bibinfo {author} {\bibfnamefont {P.}~\bibnamefont {Lochner}}, \bibinfo {author} {\bibfnamefont {F.}~\bibnamefont {Brinks}}, \bibinfo {author} {\bibfnamefont {A.}~\bibnamefont {Wieck}},\ and\ \bibinfo {author} {\bibfnamefont {A.}~\bibnamefont {Ludwig}},\ }\bibfield  {title} {\bibinfo {title} {Influence of molecular beam effusion cell quality on optical and electrical properties of quantum dots and quantum wells},\ }\href@noop {} {\bibfield  {journal} {\bibinfo  {journal} {J. Cryst. Growth}\ }\textbf {\bibinfo {volume} {550}},\ \bibinfo {pages} {125884} (\bibinfo {year} {2020})}\BibitemShut {NoStop}%
\bibitem [{\citenamefont {Chin}\ and\ \citenamefont {Luo}(1998)}]{CHIN1998}%
  \BibitemOpen
  \bibfield  {author} {\bibinfo {author} {\bibfnamefont {M.-K.}\ \bibnamefont {Chin}}\ and\ \bibinfo {author} {\bibfnamefont {C.-P.}\ \bibnamefont {Luo}},\ }\bibfield  {title} {\bibinfo {title} {Photoluminescence study of interface microroughness and exciton transfer in growth-interrupted single quantum wells},\ }\href@noop {} {\bibfield  {journal} {\bibinfo  {journal} {J. Lumin.}\ }\textbf {\bibinfo {volume} {79}},\ \bibinfo {pages} {233} (\bibinfo {year} {1998})}\BibitemShut {NoStop}%
\bibitem [{\citenamefont {Oskooi}\ \emph {et~al.}(2010)\citenamefont {Oskooi}, \citenamefont {Roundy}, \citenamefont {Ibanescu}, \citenamefont {Bermel}, \citenamefont {Joannopoulos},\ and\ \citenamefont {Johnson}}]{oskooi2010meep}%
  \BibitemOpen
  \bibfield  {author} {\bibinfo {author} {\bibfnamefont {A.~F.}\ \bibnamefont {Oskooi}}, \bibinfo {author} {\bibfnamefont {D.}~\bibnamefont {Roundy}}, \bibinfo {author} {\bibfnamefont {M.}~\bibnamefont {Ibanescu}}, \bibinfo {author} {\bibfnamefont {P.}~\bibnamefont {Bermel}}, \bibinfo {author} {\bibfnamefont {J.~D.}\ \bibnamefont {Joannopoulos}},\ and\ \bibinfo {author} {\bibfnamefont {S.~G.}\ \bibnamefont {Johnson}},\ }\bibfield  {title} {\bibinfo {title} {Meep: A flexible free-software package for electromagnetic simulations by the fdtd method},\ }\href@noop {} {\bibfield  {journal} {\bibinfo  {journal} {Comput. Phys. Commun.}\ }\textbf {\bibinfo {volume} {181}},\ \bibinfo {pages} {687} (\bibinfo {year} {2010})}\BibitemShut {NoStop}%
\end{thebibliography}
%

\section*{Acknowledgments}
We acknowledge Rodrigo A.\ Thomas for valuable discussions on the experimental methods.
We thank Alexander Vogel of the Nano Imaging Lab of the Swiss Nanoscience Institute, University of Basel for expert assistance in performing the TEM measurements.

\textbf{Funding:}
M.A.\ was funded by European Union’s Digital Europe programme (EuroQCI) under Grant agreement No.\ 101091659.
S.K., N.S., A.L.\ acknowledge BMFTR funded projects QTRAIN No.\ 13N17328, EQSOTIC No.\ 16KIS2061, and QR.N No.\ 16KIS2200, and the DFG funded project EXC ML4Q LU 2004/1.
Z.L.\ and P.L.\ acknowledge financial support from the Nordisk Foundation (Challenge project “Solid-Q”).
Y.M., P.L.\ and L.M.\ acknowledge financial support from the Danish National Research Foundation (Center of Excellence “Hy-Q,” grant No.\ DNRF139).
L.L.N., B.F.S., R.J.W.\ was funded by the Swiss National Science Foundation via Projects 20QU-1\_215955 and 200021L\-236481.
L.M.\ was funded by European Research Council (ERC) under the European Union’s Horizon 2020 research and innovation program (No.\ 949043, NANOMEQ).

\textbf{Author contributions:}
M.A., S.K., Y.M and L.M. designed and built the experimental setup and devices.
Z.L. and M.A. fabricated the sample. 
M.A., J.L. and L.S. performed the optical spectroscopy measurements. 
M.A. and L.M. carried out the data analysis. 
L.L.N. and B.F.S. carried out the transmission electron microscopy analysis.
S.K., N.S. and A.L. developed the growth protocol and performed the growth of quantum dot samples.
P.L., A.L., R.J.W. and L.M. conceptualized the idea of the O-band sample growth, the experiment, and data analysis and provided financial support for the project.  
M.A. and L.M. wrote the manuscript with contribution from all authors.

\textbf{Competing interests:}
P.~L. is founder and chief quantum officer of Sparrow Quantum. The other authors declare no competing interests.

\textbf{Data and materials availability:}
The data presented in this manuscript as well as codes used to analyze it and produce the figures are available online upon publication. The sample used is piece A3W cleaved from wafer B15835.

\end{document}


\title{ Supplementary Information for \\ A quantum-coherent photon--emitter interface in the original telecom band }

\author{
    Marcus Albrechtsen$^{1\ast\dagger}$,
    Severin Krüger$^{2\dagger}$,
    Juan Loredo$^{3}$,
    Lucio Stefan$^{3}$,
    Zhe Liu$^{1}$,
    Yu Meng$^{1}$,
    Lukas L.\ Niekamp$^{4}$,
    Bianca F.\ Seyschab$^{4}$,
    Nikolai Spitzer$^{2}$,
    Richard J.\ Warburton$^{4}$,
    Peter Lodahl$^{1,3}$,
    Arne Ludwig$^{2}$,
    Leonardo Midolo$^{1\ast}$
    \\[1ex]
    {\small \textit{
    $^{1}$Center for Hybrid Quantum Networks (Hy-Q), Niels Bohr Institute, University of Copenhagen,\\
    \hspace{1em}Jagtvej 155A, DK-2200 Copenhagen, Denmark.\\
    $^{2}$Lehrstuhl für Angewandte Festkörperphysik, Ruhr-Universität Bochum,\\
    \hspace{1em}Universitätsstraße 150, 44801 Bochum, Germany.\\
    $^{3}$Sparrow Quantum ApS, Nordre Fasanvej 215, DK-2000 Frederiksberg, Copenhagen, Denmark.\\
    $^{4}$Department of Physics, University of Basel, Klingelbergstrasse 82, CH-4056 Basel, Switzerland.}\\
    $^\ast$Corresponding authors. E-mails: m.albrechtsen@nbi.ku.dk, midolo@nbi.ku.dk\\
    $^\dagger$These authors contributed equally.
    }
}

\date{\today}

\maketitle

\renewcommand{\thefigure}{S\arabic{figure}}
\renewcommand{\thetable}{S\arabic{table}}
\renewcommand{\theequation}{S\arabic{equation}}
\renewcommand{\thepage}{S\arabic{page}}
\setcounter{figure}{0}
\setcounter{table}{0}
\setcounter{equation}{0}
\setcounter{page}{1}

\section*{Comparison with other telecom emitters}
Figure~3 compares our results to previous works in term of linewidth broadening ($\Gamma_\text{RF} / \gamma_1$) and source efficiency (defined as photon rate in the fiber divided by repetition rate of the excitation laser) for different material systems, different wavelengths in the telecom range, and different device design. Further details are provided in Table~\ref{tab:compare}. For an extended survey on the topic we also refer the reader to a recent review on telecom quantum emitters \cite{holewa2025solid}. The figure clearly shows how O-band QDs reported here, show great promise in achieving the required specifications to be employed in quantum applications that require both high coherence and loss-tolerance, such as device-independent quantum key distribution (DI-QKD) and fault-tolerant quantum computing (FTQC).

\section*{Resonant two-photon Hong-Ou-Mandel interference from a second quantum dot}
To validate our measurements of high-quality quantum emitters, we perform resonant excitation and analysis of a second quantum dot, which emit at $\lambda=\SI{1294}{nm}$ when the gate voltage is set to $V_g=\SI{-390}{mV}$. We extract a HOM visibility, $V\sim\SI{74.4}{\percent}$, when integrating across the full window but correcting for a constant background caused primarily by laser leakage. The HOM interference data is shown in Fig.~\ref{fig:SI:HOM-QD2}.

\section*{Non-resonant excitation and characterization}
Figure~\ref{fig:SI:non-resonant} shows the non-resonant excitation of a quantum dot in the gated nanostructure. Since the pump laser is detuned from the QD emission, the laser can trivially be filtered using a \SI{20}{GHz} grating filter, which results in a low $g^{(2)}(0)=1.6~\%$ (Fig.~\ref{fig:SI:non-resonant}\textbf{a}) -- still with minimal blinking.
This dot emits at $\lambda=\SI{1293.9}{nm}$ with a gate voltage $V_g=\SI{-361}{mV}$ and a lifetime, $\tau\sim\SI{310}{ps}$. Figure~\ref{fig:SI:non-resonant}\textbf{b} shows a zoom-in of the central peak, normalized to the far-away peaks. It is here evident that coincidence counts only occur within $\pm5\tau\sim\pm1.5$~ps. This dataset is not background corrected and the analysis takes into account the area across the full window $\pm6$~ns.
Figure~\ref{fig:SI:non-resonant}\textbf{c} and \textbf{d} shows the Hong-Ou-Mandel two-photon interference from this emission from both the co- and cross-polarized configuration, which is useful to understand the timescales of the decoherence processes. No background-correction or adjustments are done and raw measurements are displayed. First, all coincidence counts occur within $\pm5\tau$, which immediately imply that all counts comes from the QD emission. Second, comparing the integral under the full wavepackets yields a low visibility of $V\sim\SI{25}{\percent}$ implying that 3 out of 4 back-to-back photons are distinguishable. It can be seen that the co- and cross-polarized coincidence histograms overlaps perfectly for delays greater than a few hundred picoseconds, which implies decoherence on this timescale. This is likely caused by jitter from additional transitions required to bridge the excitation and emission energies.
Third, on short time-scales a difference manifests -- effectively illustrating that in this regime the jitter is filtered away by time-gating, and at zero time delay the HOM visibility, $V\sim\SI{87}{\percent}$ approaches the Hanbury Brown and Twiss $g^{(2)}(0)$, only it is broadened by the instrument response function (i.e., the joint jitter of the two SNSPD and TimeTagger channels). In other words, the post-selected indistinguishable photons are more indistinguishable, however, such filtering comes at a hefty price in terms of efficiency, and a post-selected HOM visibility should not be misinterpreted as a sign of coherent emission limited only by pure dephasing.

\section*{Characterization and documentation of the optical setup}
Figure~\ref{fig:sup_setup} shows a schematic of the optical setup,
where M1--8 is a broadband dielectric mirror (BB1-EO4), 1310PMY is a panda 1310 polarization maintaining fiber, TC06APC-1310 (TC25APC-1310) represent fixed-focus triplet collimators with beam waists diameter of 1.11~mm and 4.58~mm, respectively, at a 1310~nm wavelength.
PBS is a polarizing beam splitter (PBS253), LP$_{(1,2)}$ are linear polarizers (LPNIR and LPNIRD050, respectively), $\lambda/2$ and $\lambda/4$ are achromatic half- and quarter-wave plates (Foktek Photonics, Inc., $T>\SI{99.8}{\percent}$), L1--3(4) are achromatic lens doublet with focal length $f=300$mm (100mm) forming a 4$f$ system between the objective (OBJ, attocube LT-APO/IR/0.81) and the input/output TC06APC collimator as well as to M1 and M4. The objective has $\text{NA}=0.80$ with a clear aperture $\SI{4.7}{mm}$ and a transmission $T\sim\SI{75}{\percent}$ measured at room temperature.
BS 50:50 are non-polarizing 50:50 beamsplitter cubes (BS015), and BS$_z$ 50:50 is a non-polarizing 50:50 beamsplitter cube (BS012) on a translation stage, which allows it to be moved in and out of the collection path.
PM: Germanium photodiode (S122C) linked with PID-control of the laser. LED: Light-emitting diode (M1300L4). The BS for the LED-path is on a flip-mount. CAM: visible and infrared camera (Atlas IP67 1G IMx991, 0.3~MP), Etalon: Fused Silica etalon with 1.5~GHz bandpass and $T\sim25$~\% (OP-13104-1686-6, LightMachinery).
DMLP1000R: Dichroic mirror that allows access for an above-band laser and allows the setup to be used for both telecom and NIR applications. 4K cryo: AttoDry800 cryostat at 4.0--4.3~K.
Unless another brand or dimension is specified, components are 1'' and from Thorlabs. The setup beam-height is 6.35~mm, except M5, M6 and DMLP1000R and L2, which forms a periscope to reach into the cryostat through the top-view anti-reflection coated window.
Crucially, the input and output paths are separated by a beamsplitter (BS$_z$) on a translation stage in a 4$f$ configuration, which allows the collection path to go through it during alignment, and then, by translating this beamsplitter, the collection path goes above or below it while the input path remains through it.

The insets shows two reference circuits, identical except one is with and the other without a photonic crystal waveguide, which allows estimation of the insertion loss. Figure~\ref{fig:sup_trans_fgc_phc} shows transmission spectra through two such reference circuits using a supercontinuum laser and a spectrometer. The fringes at longer wavelengths originate from the focusing grating coupler \cite{zhou2018high}, as the grating is slightly blue-shifted from its optimal design due to shallow-etched features not being etched deep enough. The circuit with the PhC waveguide clearly shows the presence of the bandedge around 1304~nm, above which no transmission is allowed.
The main QD resonance is indicated with a vertical line, as is the wavelength 1290~nm, which is used to calibrate the transmission in absolute numbers and this result is reported in Table~\ref{table:efficiency}.

Figure~\ref{fig:sup_propLoss}\textbf{a} shows a scanning electron microscope image of 3 spiral waveguide circuits \cite{wang2021electroabsorption}, identical except for the length of the arms thereby enabling a cutback measurement to separate the propagation loss in the waveguides from the total loss.
Figure~\ref{fig:sup_propLoss}\textbf{b} shows transmission spectra through circuits of 7 different lengths in a chip from a different wafer without a $p$-$i$-$n$ diode. The grating coupler were not blue-shifted on this sample, thus having it peak around the target 1300~nm wavelength, however, fringes appear due to  spurious reflections in the device and to catch the propagation loss accurately, we consider the slope with the positive envelope of the data \cite{ljubotina2025origins} shown in Fig.~\ref{fig:sup_propLoss}\textbf{c} and indicating a mean propagation loss of 1.5~dB/mm.
We also carry out this analysis on our gated membrane, whose stack is shown in Fig.~\ref{fig:sup_propLoss}\textbf{d} and where the $p$-doped layer leads to additional absorptive losses. The cutback data is shown in Fig.~\ref{fig:sup_propLoss}\textbf{e} and the slope in Fig.~\ref{fig:sup_propLoss}\textbf{f}, which results in a mean propagation loss of 3.5~dB/mm.
We discuss origins and mitigation of the propagation losses in our suspended waveguides in more detail in Ref.~\cite{ljubotina2025origins}, and note that excellent propagation loss in GaAs have been demonstrated at telecom wavelengths, far from the bandgap \cite{ottaviano2016,chang2020}.

\begin{figure*}
	\centering
	\includegraphics[width=0.8\textwidth]{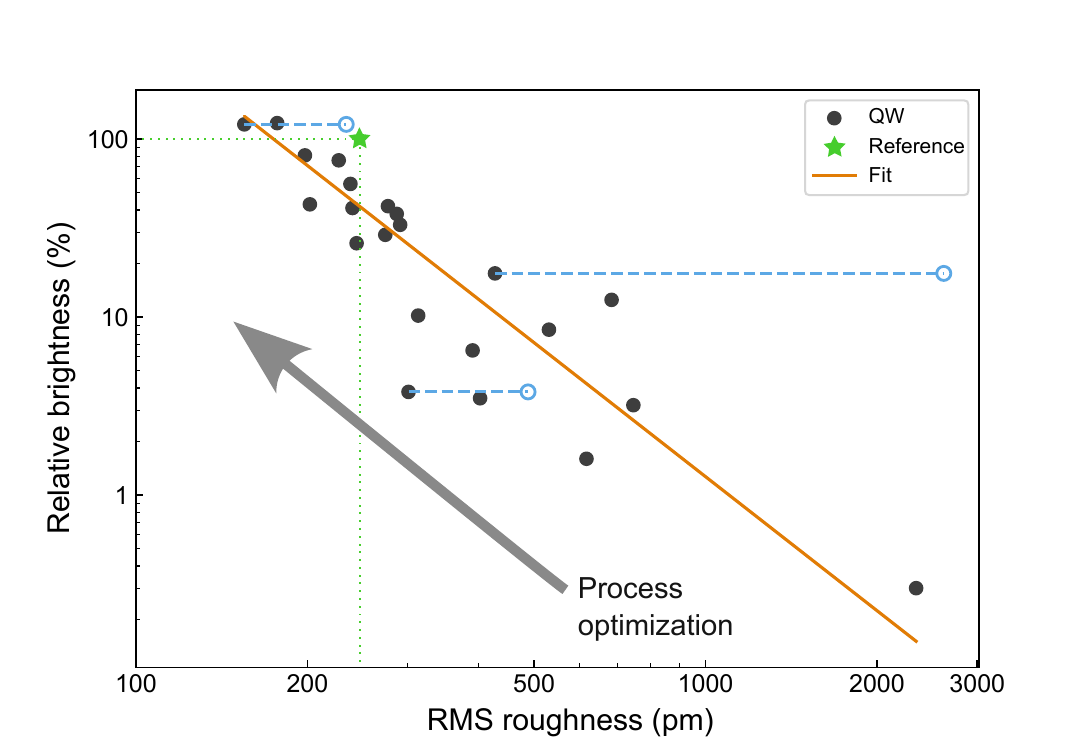}
	\caption{\textbf{Quantum well brightness correlated to roughness.} Each data point (black dot) corresponds to a test quantum well (QW) sample grown during the process development and optimization. The QW brightness is normalized to a high-quality reference sample \cite{NGUYEN2020} (green star). The fit was done with a power function $ax^{b}$, yielding $a=4.04\cdot10^7$ and $b=-2.50$. Dashed lines indicate the increase in roughness compared to the measured QW sample, when growing the full optical stack including DBR layers and membrane (blue circles).}
	\label{fig:correlation}
\end{figure*}

\begin{figure*}
	\centering
	\includegraphics[width=0.8\textwidth]{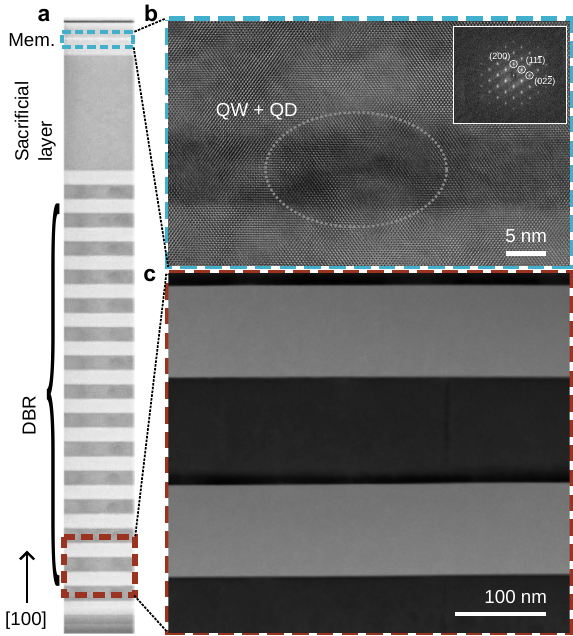}
	\caption{\textbf{Scanning transmission electron microscope analysis of the sample.} \textbf{a}, Full stack including the distributed Bragg reflector (DBR) and membrane with QDs. \textbf{b}, High-resolution scan of the QW and QD area. The circled line indicates the rough position of QDs. No visible defects are found. Inset: Fast Fourier transform of the image. Indexing is shown for zone axis $[0\bar{1}\bar{1}]$. \textbf{c}, Detail of the DBR region, showing sharp transitions between layers.}
	\label{fig:DBR}
\end{figure*}

\begin{figure*}
    \centering
    \includegraphics[width=0.6\textwidth]{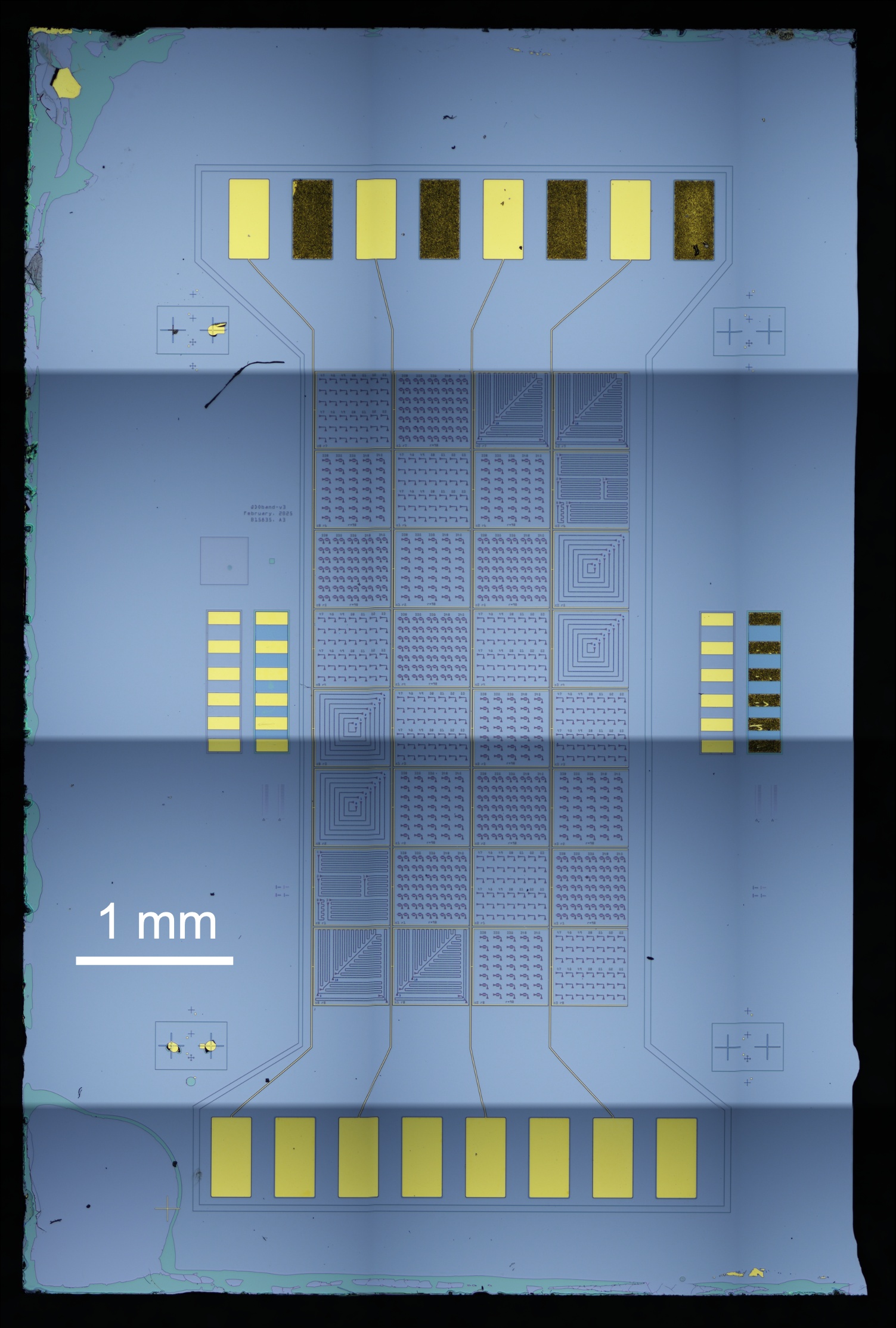}
    \caption{
        \textbf{Stitched optical microscope-image (5x) of the chip.} The device used is in the rightmost column (4), second row from the bottom. The MESA for each column is $\sim\SI{2.5}{\mm\squared}$. The transmission-line-mesas (TLMs) on either side is used to characterize sheet and contact resistances to verify Ohmic behaviour for $p$-$p$ and $n$-$n$ and a diode across the $p$-$i$-$n$ junction. Dark bonding pads are annealed ohmic $n$-contacts (used). Bright bonding pads are ohmic $p$-contacts (used) or Schottky n-contacts (not used).
    }
    \label{fig:sup_microscope}
\end{figure*}

\begin{figure*}
    \centering
    \includegraphics[width=0.9\textwidth]{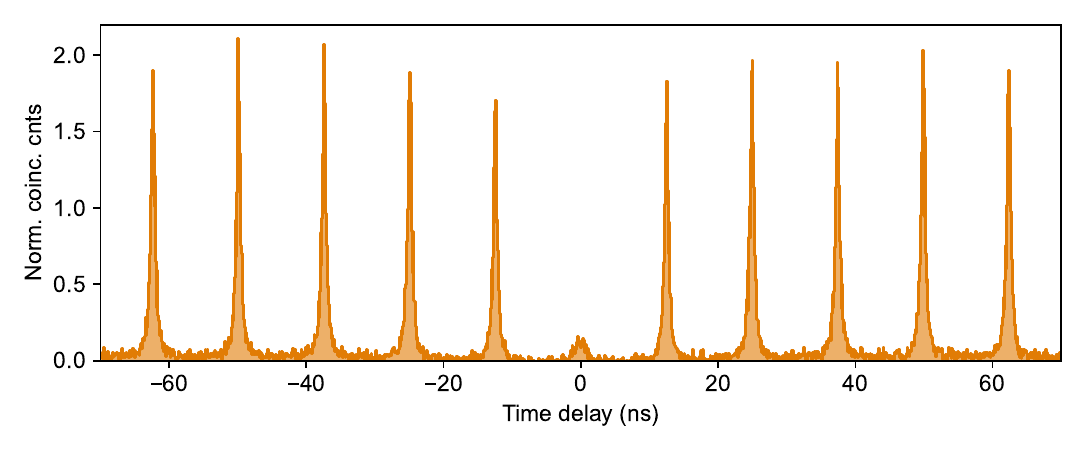}
    \caption{
        \textbf{Hong-Ou-Mandel (HOM) interference measurement of a second quantum dot.}
        The quantum dot is excited resonantly with the same pulsed laser pulsed laser at gate voltage, $V=\SI{-390}{mV}$, and has lifetime, $\tau\sim\SI{450}{ps}$.
        This measurement is done without the narrow etalon but using a \SI{20}{GHz} grating filter, and is corrected for a constant background. The resulting HOM visibility (integrating across the full \SI{12}{ns} window) is $\sim\SI{74.4}{\percent}$.
        }
    \label{fig:SI:HOM-QD2}
\end{figure*}

\begin{figure*}
    \centering
    \includegraphics[width=0.9\textwidth]{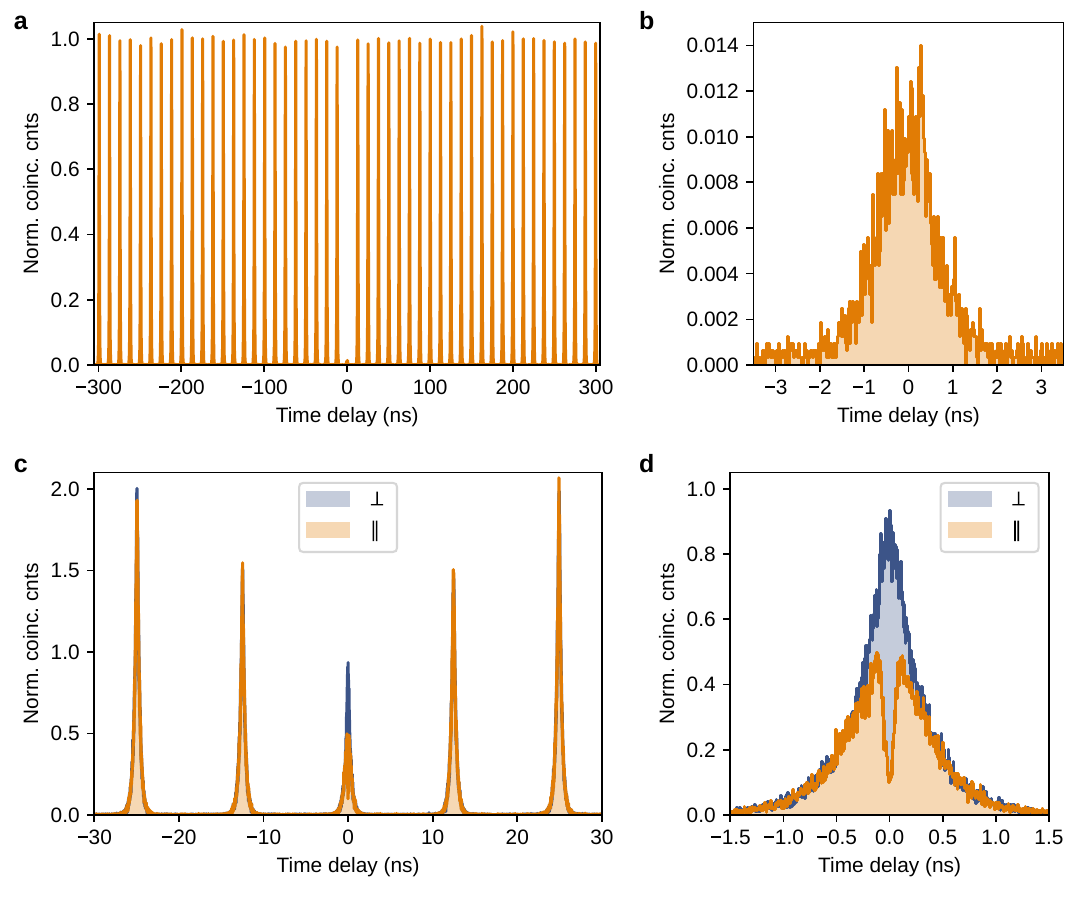}
    \caption{
        \textbf{Non-resonant excitation.}
        A quantum dot with $\tau\sim\SI{310}{ps}$ and emission $\lambda\sim\SI{1293.9}{nm}$ at gate voltage, $V\sim\SI{-361}{mV}$ is excited non-resonantly, which allows suppression of the excitation laser. \textbf{a}, Hanbury Brown and Twiss (HBT) correlation measurement with a zoom-in of the central peak in \textbf{b}, evidencing pure single-photon emission with $g^{(2)}(0)=\SI{1.6}{\percent}$.
        \textbf{c}--\textbf{d}, Hong-Ou-Mandel (HOM) interference of the single photons illustrating a degraded coherence across the full wave packet ($V=\SI{25}{\percent}$) due to jitter from the non-resonant process compared to the coherence achieved at zero time-delay only ($V=\SI{87}{\percent}$ at dip).
        All data is raw, i.e., no background subtraction.
        }
    \label{fig:SI:non-resonant}
\end{figure*}

\begin{figure*}
    \centering
    \includegraphics[width=0.6\textwidth]{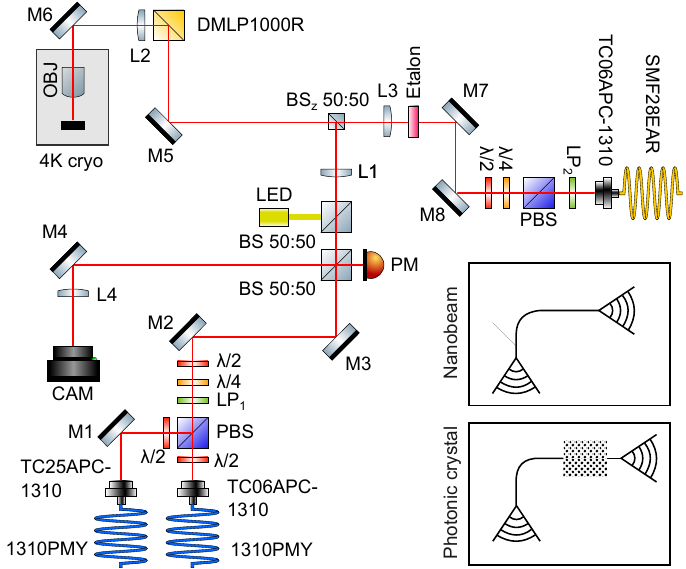}
    \caption{
        \textbf{Schematic of the optical setup.}
        The insets indicates two on-chip calibration circuits, with and without a photonic crystal waveguide, to isolate the spectral insertion loss of this component.
    }
    \label{fig:sup_setup}
\end{figure*}

\begin{figure*}
    \centering
    \includegraphics[width=0.6\textwidth]{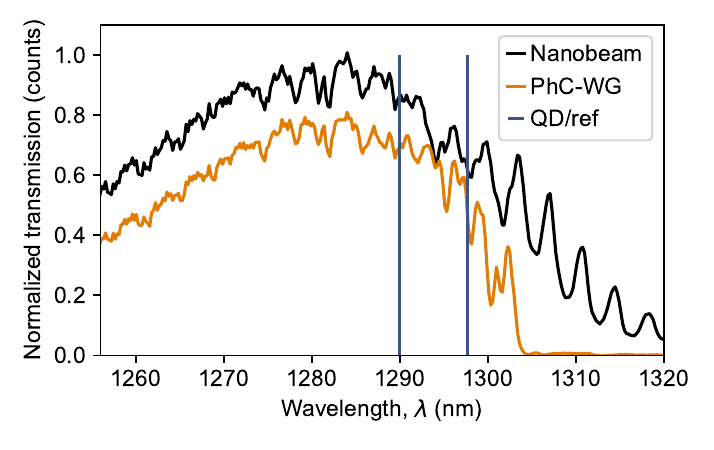}
    \caption{
        \textbf{Collection efficiency including photonic crystal and grating coupler performance.}
        Transmission spectrum through a suspended nanobeam waveguide revealing the isolated performance of the focusing rating coupler (FGC) and the transmission through the same device with coupling to and from nanobeam-waveguide to photonic crystal (PhC) waveguide ($a=\SI{338}{nm}$, $r=\SI{90}{nm}$). The quantum dot emission wavelength is indicated as is the \SI{1290}{nm} wavelength at which the setup transmission is measured with a continuous wave laser.
    }
    \label{fig:sup_trans_fgc_phc}
\end{figure*}

\begin{figure*}
    \centering
    \includegraphics[width=0.9\textwidth]{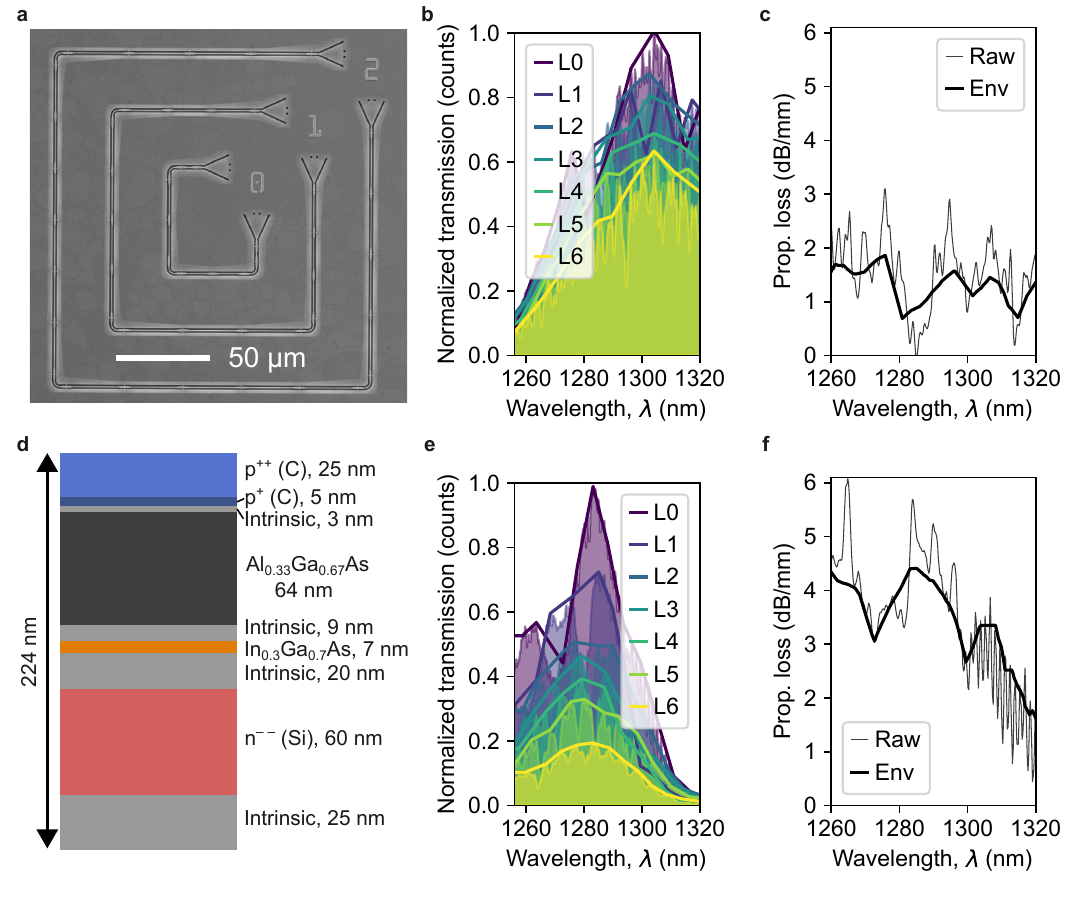}
    \caption{
        \textbf{Cutback characterization of the propagation loss in the planar GaAs waveguides.}
        \textbf{a}, Scanning electron microscope (SEM) image showing 3 of 7 circuits whose only difference is the length of the straight-line sections. The lengths are: (120, 360, 600, 840, 1080, 1320, 1560)~\SI{}{\micro\m}. The waveguides are $\sim\SI{620}{\nm}$ wide.
        \textbf{b}, Transmission through the 7 circuits on an undoped reference chip (b15641b6w), reflections causes spurious resonances resulting in random transmission dips and therefore we also compute the envelopes. \textbf{c}, Propagation loss (i.e., slope to a linear fit at each wavelength) considering both the raw data ($(1.5\pm0.6)~\SI{}{\dB/\mm}$) and the envelope ($(1.3\pm0.3)~\SI{}{\dB/\mm}$).
        \textbf{d}, Material stack of the $p$-$i$-$n$ gated quantum dots, with additional loss due to free-carrier absorption primarily in the $p$-layer with transmission shown in \textbf{e} (incl. a spectral blue-shift due to fabrication variations) and the measured propagation loss for the raw ($(3.5\pm1.2)~\SI{}{\dB/\mm}$) and envelope ($(3.5\pm0.7)~\SI{}{\dB/\mm}$) data shown in \textbf{f}. Losses are stated as mean~$\pm$~standard deviation in the displayed range.
    }
    \label{fig:sup_propLoss}
\end{figure*}


\begin{table*}
	\centering
	\caption{ \label{tab:si_structure} Full layer stack of the device grown by molecular beam epitaxy.}
	\begin{tabular}{|l|l|l|l|c|}
	\hline
	Material & Thickness [\si{\nano\meter}] & Growth T [\si{\degreeCelsius}]& Loop & Purpose \\\hline
	GaAs:C & 25 & 600 && p-doping (\num{1.2e19}~\SI{}{\per\cm\cubed}) \\\hline
	GaAs:C & 5 & 600 && p-doping (\num{2.5e18}~\SI{}{\per\cm\cubed}) \\\hline
	GaAs & 3 & 600 & & \\\hline
	$\mathrm{Al}_{0.33}\mathrm{Ga}_{0.67}\mathrm{As}$ & 64 & 600 && barrier \\\hline
	GaAs & 9 & 445 & & \\\hline
	$\mathrm{In}_{0.3}\mathrm{Ga}_{0.7}\mathrm{As}$ & 7 & 445 & & strain reduction layer \\\hline
	InAs & 0.5 & 495 & & quantum dots \\\hline
	GaAs & 20 & 600 && tunnel barrier and PDL \\\hline
	GaAs:Si & 60 & 600 && n-doping (\num{2.0e18}~\SI{}{\per\cm\cubed}) \\\hline
	GaAs & 25.6 & 600 && \\\hline
	$\mathrm{Al}_{0.76}\mathrm{Ga}_{0.24}\mathrm{As}$ & 865 & 600 && sacrifical layer \\\hline
	GaAs & 97 & 600 &15x& AlAs/GaAs DBR \\
	AlAs & 112 & 600 & &  \\\hline
	GaAs & 94.4 & 600 & & \\\hline
	GaAs & 2.6 & 600 & 14x & superlattice \\
	AlAs & 2.6 & 600 & & \\\hline
	GaAs & 200 & 600 && buffer \\\hline
	GaAs & 0.625 mm & - && substrate \\\hline
	\end{tabular}
\end{table*}

\begin{table*}
\centering
\caption{\textbf{State-of-the-art of telecom single-photon emitters and their main properties.} PCW: Photonic Crystal Waveguide, FC: Frequency conversion, DBR: Distributed Bragg Reflector, CBG: Circular Bragg Grating, $\mu$MESA: Micro-mesa, PCC: Photonic Crystal Cavity. $V_\text{HOM}$ used here sometimes refers to two-photon interference (TPI) visibility in other works. The table only provides an overview, we refer the reader to the specific works for precise details (such as uncertainties) related to the specific experiments. }
\label{tab:compare}
	
\begin{tabular}{r|lllllll}
\hline
Ref. & Device & $\lambda$ [$\mu$m] & $g^2$(0) [\%]  & $V_\text{HOM}$ & $\tau$ [ns] & Counts [MHz] & Pump [MHz]\\
\hline
\hline
\textcolor[rgb]{0.121,0.465,0.703}{$\star$} This work & PCW & 1.300 &  6.1 & 0.838 & 0.150 & 7.6 &  80.192\\
\textcolor[rgb]{0.996,0.496,0.055}{$\diamond$} \cite{zahidy2024quantum} & PCW+FC & 1.545 &  0.47 & 0.91 & 0.867 & 6 & 72.6\\
\textcolor[rgb]{0.172,0.625,0.172}{$\lhd$} \cite{nawrath2019coherence} & DBR & 1.554 &  2.3 & 0.713 & 1.71 & 0.016 & 82.86\\
\textcolor[rgb]{0.836,0.152,0.156}{$\rhd$} \cite{hauser2025deterministic} & CBG & 1.550 &  1.7 & 0.917 & 0.257 & 0.43 &  80\\
\textcolor[rgb]{0.578,0.402,0.738}{$\blacksquare$} \cite{joos2024coherently} & CBG & 1.555 &  7.6 & 0.104 & 0.46 & 8.1 & 152\\
\textcolor[rgb]{0.547,0.336,0.293}{$\nabla$} \cite{nawrath2023bright} & CBG & 1.555 &  0.72 & 0.081 & 1.59 & 14 & 228\\
\textcolor[rgb]{0.887,0.465,0.758}{$\oplus$} \cite{holewa2024high} & CBG & 1.554 &  0.47 & 0.193 & 0.4 & 0.11 &  80\\
\textcolor[rgb]{0.496,0.496,0.496}{$\otimes$} \cite{srocka2020deterministically} & $\mu$MESA & 1.300 &  2.7 & 0.124 & 1.6 & 0.17 &  80\\
\textcolor[rgb]{0.734,0.738,0.133}{$\ast$} \cite{komza2024indistinguishable} & PCC & 1.280 &  15 & 0.004 & 4.6 & 0.018 &  40\\
\textcolor[rgb]{0.090,0.742,0.809}{$\vee$} \cite{ourari2023indistinguishable} & PCC & 1.550 &  1.8 & 0.6 & 7.4e+03 & 0.2 & 5.7\\
\hline
\end{tabular}
\end{table*}

\begin{table*}
	\centering
	\caption{
            \textbf{System efficiency and setup losses.}
            Losses from individual components are isolated. The PhC-nanobeam and the grating coupler losses are estimated from Fig.~\ref{fig:sup_trans_fgc_phc}, the objective transmission is measured at \SI{300}{K}. Fiber coupling is measured using a pin-hole in the 4F-system, the grating filter and the free-space optics are measured. The SNSPDs is taken from the data-sheet, which specifies \SI{72}{\percent} at 5 MHz count rates at \SI{1310}{nm} and slightly lower at \SI{1297.7}{nm}. Losses from fiber-connections and splices are not included and assumed to be unity.
            The efficiency from QD to detector is \SI{8.4}{\percent}, from QD to fiber-coupling is \SI{16.0}{\percent}, from QD to first lens is \SI{35.7}{\percent}, and from QD to nanobeam waveguide is \SI{87.8}{\percent}.
                }
	\label{table:efficiency}
	
	\begin{tabular}{lccc}
		\\
		\hline
		Description & Transmission (\%) \\
		\hline
            PhC $\rightarrow$ nanobeam-WG & \SI{87.8}{\percent} \\
            Grating coupler & \SI{40.6}{\percent} \\
            Objective & \SI{75.0}{\percent} \\
            Free-space optics & \SI{92.8}{\percent} \\
            Fiber-coupling & \SI{64.5}{\percent} \\
            Grating filter & \SI{75.0}{\percent} \\
            SNSPDs & \SI{70.0}{\percent} \\
		\hline
	\end{tabular}
\end{table*}